\documentclass[a4paper,11pt]{article}

\usepackage{amsmath,amssymb}
\usepackage{graphicx}

\makeatletter

 \@addtoreset{equation}{section}
\makeatother

\newcommand{\vev}[1]{\langle {#1} \rangle}
\newcommand{\del}{\partial}
\newcommand{\half}{\frac{1}{2}}
\newcommand{\mbf}{\boldsymbol}
\newcommand{\LS}{\ \ \ \ \ \ \ \ \ \ }
\newcommand{\ls}{\ \ \ \ \ }
\newcommand{\wt}{\widetilde}
\newcommand{\wh}{\widehat}
\newcommand{\ve}{\varepsilon}
\newcommand{\ol}{\overline}

\newcommand{\bsubeq}{\begin{subequations}}
\newcommand{\esubeq}{\end{subequations}}
\newcommand{\noi}{\noindent}
\renewcommand{\d}{{\rm d}}
\newcommand{\nn}{\nonumber}
\newcommand{\C}{{\mathbb C}}

\newcommand{\CP}[2]{{\bf W}\C{\bf P}^{#1}_{#2}}

\renewcommand{\P}[1]{\C{\bf P}^{#1}}
\newcommand{\e}{{\rm e}}
\newcommand{\all}{{}^{\forall}}
\newcommand{\exist}{{}^{\exists}}
\DeclareFontFamily{U}{rsf}{}
\DeclareFontShape{U}{rsf}{m}{n}{
  <5> <6> rsfs5 <7> <8> <9> rsfs7 <10-> rsfs10}{}
\DeclareMathAlphabet\Scr{U}{rsf}{m}{n}
\makeatletter
\def\hhline{%
  \noalign{\ifnum0=`}\fi\hrule\@height3\arrayrulewidth \futurelet
   \reserved@a\@xhline}
\makeatother
\makeatletter
\def\hvline{\vrule \@width 3\arrayrulewidth}
\makeatother
\makeatletter
\def\fhline{%
  \noalign{\ifnum0=`}\fi\hrule\@height.5\arrayrulewidth \futurelet
  \reserved@a\@xhline}
\makeatother
\makeatletter
\def\fvline{\vrule \@width.5\arrayrulewidth}
\makeatother
\makeatletter
\def\hcline#1{\@hcline#1\@nil}
\def\@hcline#1-#2\@nil{%
  \omit
  \@multicnt#1%
  \advance\@multispan\m@ne
  \ifnum\@multicnt=\@ne\@firstofone{&\omit}\fi
  \@multicnt#2%
  \advance\@multicnt-#1%
  \advance\@multispan\@ne
  \leaders\hrule\@height3\arrayrulewidth\hfill
  \cr
  \noalign{\vskip-\arrayrulewidth}}
\def\multispan{\omit\@multispan}
\def\@multispan#1{%
  \@multicnt#1\relax
  \loop\ifnum\@multicnt>\@ne \sp@n\repeat}
\def\sp@n{\span\omit\advance\@multicnt\m@ne}
\makeatother
\makeatletter
\def\fcline#1{\@fcline#1\@nil}
\def\@fcline#1-#2\@nil{%
  \omit
  \@multicnt#1%
  \advance\@multispan\m@ne
  \ifnum\@multicnt=\@ne\@firstofone{&\omit}\fi
  \@multicnt#2%
  \advance\@multicnt-#1%
  \advance\@multispan\@ne
  \leaders\hrule\@height.5\arrayrulewidth\hfill
  \cr
  \noalign{\vskip-\arrayrulewidth}}
\def\multispan{\omit\@multispan}
\def\@multispan#1{%
  \@multicnt#1\relax
  \loop\ifnum\@multicnt>\@ne \sp@n\repeat}
\def\sp@n{\span\omit\advance\@multicnt\m@ne}
\makeatother
\makeatletter
\def\bcline#1{\@bcline#1\@nil}
\def\@bcline#1-#2\@nil{%
  \omit
  \@multicnt#1%
  \advance\@multispan\m@ne
  \ifnum\@multicnt=\@ne\@firstofone{&\omit}\fi
  \@multicnt#2%
  \advance\@multicnt-#1%
  \advance\@multispan\@ne
  \cleaders\hbox{$\m@th \mbox{\rule{.1em}{.035em}}\mkern2mu$}\hfill
  \cr
  \noalign{\vskip-\arrayrulewidth}}
\def\multispan{\omit\@multispan}
\def\@multispan#1{%
  \@multicnt#1\relax
  \loop\ifnum\@multicnt>\@ne \sp@n\repeat}
\def\sp@n{\span\omit\advance\@multicnt\m@ne}
\makeatother

\begin{document}
\allowdisplaybreaks{

\setcounter{page}{0}

\begin{titlepage}

{\normalsize
\begin{flushright}
{\tt KEK-TH-999}\\ 
{\tt hep-th/0411258}\\
November 2004
\end{flushright}
}

\vspace{7mm}

\begin{center}
{\Huge Gauged Linear Sigma Models

\vspace{5mm}

for Noncompact Calabi-Yau Varieties}

\vspace{10mm}

\bigskip
{\Large Tetsuji {\sc Kimura}}

\vspace{3.5mm}

{\sl
Theory Division, Institute of Particle and Nuclear Studies,\\
High Energy Accelerator Research Organization (KEK)\\
Tsukuba, Ibaraki 305-0801, Japan

\vspace{1mm}

\tt tetsuji@post.kek.jp}

\end{center}

\vspace{8mm}


\begin{abstract}

We study gauged linear sigma models for noncompact Calabi-Yau
manifolds described as a line bundle on a hypersurface in a projective space.
This gauge theory has a unique phase if the Fayet-Iliopoulos parameter 
is positive, 
while there exist two distinct phases if the parameter is negative.
We find four massless effective theories in the
infrared limit,
which are related to each other under 
the Calabi-Yau/Landau-Ginzburg correspondence and the topology change.
In the T-dual theory, on the other hand,
we obtain two types of exact massless effective
theories:
One is the sigma model on a newly obtained Calabi-Yau geometry as
a mirror dual,
while the other is given by a Landau-Ginzburg theory with a negative
power term,
indicating ${\cal N}=2$ superconformal field
theory on $SL(2,{\mathbb R})/U(1)$.
We argue that the effective theories in the original gauged
linear sigma model are exactly realized as ${\cal N}=2$
Liouville theories coupled to well-defined Landau-Ginzburg minimal models.

\end{abstract}

\end{titlepage}

\newpage
\setcounter{page}{2}

\section{Introduction}

Field theory in two-dimensional spacetime is one of the most
powerful tools for analyzing dynamical phenomena in particle physics.
It has been studied as a toy model of low energy effective theory
including symmetry breaking mechanism.
Nonlinear sigma model (NLSM) on the projective space is a typical example to
investigate chiral symmetry breaking \cite{diV79, GN}.
String worldsheet theory is also described as a two-dimensional conformal
field theory (CFT).
Conformal invariance in the worldsheet theory gives a set of equations of
motion of spacetime physics \cite{FMS}.
When we discuss string theory on curved spacetime, 
supersymmetric NLSMs play significant roles.

Coupled to a gauge field, two-dimensional field theory
has been applied to more complicated physics.
The gauge field plays a key role 
in the compactification of the target space via the Higgs mechanism.
This theory is also useful to study mathematical problems
such as Morse theory \cite{W88} and mirror symmetry \cite{GP90}.
Many people have constructed two-dimensional supersymmetric
gauge theories in order to understand various kinds of physical and
mathematical structures.

Higashijima and Nitta formulated
supersymmetric NLSMs on hermitian symmetric spaces (HSSs),
which are specific K\"{a}hler manifolds,
starting from supersymmetric gauge theories with four supercharges \cite{HN99}.
By using this, 
we constructed K\"{a}hler metrics on 
complex line bundles over compact Einstein-K\"{a}hler manifolds
\cite{HKN0110, HKN0202}.
These noncompact K\"{a}hler manifolds have vanishing Ricci tensors,
and hence are Calabi-Yau (CY) manifolds \cite{Calabi, Yau}.
In attempt to investigate the gauge/gravity duality in string
theory \cite{KW98, KS01, CGLP01} on these CY manifolds, however, 
it is indispensable to understand
 global aspects such as cohomology
classes.

On the other hand, 
it is a well-known fact that 
the gauged linear sigma model (GLSM) is useful
to investigate worldsheet string theories on toric varieties \cite{W93}. 
In this framework,
we can understand some global properties of the CY manifolds.
GLSM includes at least two kinds of SCFTs in the infrared (IR) limit.
One is a supersymmetric NLSM on CY manifold
and the other is an ${\cal N}=2$ Landau-Ginzburg (LG) theory.
We can read the cohomology ring of the CY manifold from 
the chiral ring derived from the LG superpotential \cite{LVW89}.
Furthermore, 
its T-dual theory provides us with mirror descriptions 
of the original geometry \cite{HV00, Mirror03}.

Since
each HSS is constructed as a submanifold of a complex projective space
or of a Grassmannian, 
we can, in general, construct the GLSMs for the line bundles on HSSs.
Investigating LG theories in the IR limit of the GLSMs,
we will be able to understand cohomology rings of the noncompact CY
manifolds.
Unfortunately, however, 
it is difficult to embed the sigma models on HSSs 
into the GLSMs.
Thus we study the GLSM for a line bundle of homogeneous hypersurface
in the projective space whose Ricci tensor vanishes \cite{TK0409}.
This noncompact manifold is represented as 
${\cal O}(-N+\ell)$ bundle on $\P{N-1}[\ell]$, 
which can be seen as a toy model of the line bundles on HSSs.

In this paper we will find the following theories in the IR limit:
NLSMs on CY manifolds,
orbifolded LG theories, 
gauged Wess-Zumino-Witten (WZW) models 
on coset $SL(2,{\mathbb R})/U(1)$ and Liouville theories.
They appear as ${\cal N}=2$ SCFTs.
The former two theories give unitary conformal field theory.
Under the conformal invariance, 
the sigma model and the LG theory become appropriate SCFTs 
from differential geometric and algebro-geometric points of view, respectively.
They often emerge when we analyze superstring theory on compact
manifolds \cite{G871, G872}.
The latter two theories are slightly different.
These theories appear in string theory on noncompact
curved spacetime such as a two-dimensional black hole \cite{W91WZW, HK01}.
They are also utilized in non-critical string theory and matrix model 
\cite{GM, NaM}.
It is quite important to study all four SCFTs simultaneously
when we consider string theories on noncompact CY manifolds.
Thus we study
the GLSM for noncompact CY manifolds including 
all the above four theories in the low energy limit.

This paper is organized as follows.
In section \ref{O-CPNk}
we study the GLSM for line bundles
and discuss how massless effective theories
appear in some specific limits.
In this analysis we find that 
there exist two distinct phases in the negative FI parameter region.
This phenomenon newly appears, while other well-known GLSMs do not
include this.
In section \ref{HV-O-CPNk} we discuss the T-dual of the GLSM.
We obtain two types of exact effective theories,
the sigma models on newly constructed mirror CY geometries and 
the LG theories with negative power.
There we discuss the exact effective theories in the original GLSM. 
We devote section \ref{summary} to the summary and discussions.
In appendix \ref{convention}
we introduce conventions of ${\cal N}=2$ supersymmetry 
in two-dimensional spacetime.
In appendix \ref{wcp}
we review a definition of weighted projective space. 
Finally we briefly introduce the linear dilaton CFT and 
discuss an interpretation of LG superpotential with
a negative power term in appendix \ref{Giveon-Kutasov}.
This argument is useful to understand the LG theories
in section \ref{HV-O-CPNk}.

\section{Gauged linear sigma model} \label{O-CPNk}

\subsection{Lagrangian: review} \label{GLSM-conv}

First of all,
let us briefly review of a general formulation of 
the GLSM \cite{W93}.
In this model there appear various superfields 
such as a chiral superfield $\Phi_a$, a vector
superfield $V$ and a twisted chiral superfield $\Sigma$,
whose definitions are in appendix \ref{convention}. 
We also incorporate a complexified abelian gauge transformation
\begin{gather*}
\Phi_a \ \to \ \Phi_a' \ = \ \e^{ -2i Q_a \Lambda } \, \Phi_a \; , \ls 
\ol{\Phi}{}_a \ \to \ \ol{\Phi}{}_a' \ = \ 
\e^{ + 2 i Q_a \ol{\Lambda} } \, \ol{\Phi}{}_a 
\; , \\
V \ \to \ V' \ = \ V + i (\Lambda - \ol{\Lambda}) \; , \ls
\Sigma \ \to \ \Sigma' \ = \ \Sigma \; ,
\end{gather*}
where $Q_a$ is a $U(1)$ charge of the chiral superfield $\Phi_a$.
For convenience, we restrict these charges to integers: 
$Q_a \in {\mathbb Z}$.
The complexified gauge parameters are described by 
a chiral and an anti-chiral superfields $\Lambda (x, \theta, \ol{\theta})$
and 
$\ol{\Lambda} (x, \theta, \ol{\theta})$, respectively:
$\ol{D}{}_{\pm} \Lambda = 0$, $D_{\pm} \ol{\Lambda} = 0$.
By using superfields,
we construct a supersymmetric gauge invariant Lagrangian:
\begin{align*}
{\Scr{L}}_{\rm GLSM} \ &= \ 
\int \! \d^4 \theta \, \Big\{ 
- \frac{1}{e^2} \ol{\Sigma} \Sigma
+ \sum_a \ol{\Phi}{}_a \, \e^{2 Q_a V} \Phi_a
\Big\} 
\\
\ & \ \ \ \ 
+ \Big( \frac{1}{\sqrt{2}} 
\int \! \d^2 \wt{\theta} \, \wt{W} (\Sigma) + c.c. \Big) 
+ \Big( \int \! \d^2 \theta \, W_{\rm GLSM} (\Phi_a) + c.c. \Big) 
\; , 
\end{align*}
where 
we assume that all chiral superfields have non-zero $U(1)$ charges
$Q_a \neq 0$ because a neutral chiral superfield is completely free
from the system. 
The abelian gauge coupling constant $e$, which appears in front of the
kinetic term of $\Sigma$, has mass dimension one.
There exist two types of superpotentials. 
One is a superpotential written as $W_{\rm GLSM} (\Phi_a)$. 
This is a holomorphic function of chiral superfields $\Phi_a$.
The other is called a twisted superpotential $\wt{W}(\Sigma)$ described as
\begin{align*}
\wt{W}(\Sigma) \ &= \ - \Sigma \, t \; , \ \ \ 
t \ = \ r -i \theta \; ,
\end{align*}
where $t$ is a complex parameter defined by the Fayet-Iliopoulos 
(FI) parameter $r$ and 
the Theta-angle $\theta$.
We also refer $t$ to the (complexified) FI parameter.

We are interested in supersymmetric low energy effective theories.
Thus we need to study the potential energy density ${\cal U} (\varphi)$
described by the scalar component fields of superfields:
\bsubeq \label{SUSY-fn}
\begin{gather}
{\cal U} (\varphi) \ = \ 
\frac{e^2}{2} {\cal D}^2 
+ \sum_a \big| F_a \big|^2 
+ {\cal U}_{\sigma} (\varphi)
\; , \label{pot-0} \\
{\cal D} \ := \ \frac{1}{e^2} D \ = \ r - \sum_a Q_a |\phi_a|^2 
\; , \ \ \ 
\ol{F}_a \ = \ - \frac{\del}{\del \phi_a} W_{\rm GLSM} (\phi) 
\; , \ \ \
{\cal U}_{\sigma} (\varphi) \ := \ 2 |\sigma|^2 \sum_a Q_a^2 |\phi_a|^2
\; . \label{D-cond} 
\end{gather}
\esubeq
Where the scalar components of $\Phi_a$ and $\Sigma$ are expressed
by $\phi_a$ and $\sigma$.
We sometimes abbreviate scalar component fields of all
superfields to $\varphi_a$.
The functions $D$ and $F_a$ 
are auxiliary fields of $\Sigma$ and $\Phi_a$, respectively. 
We need not include fermionic components into the above functions 
(\ref{SUSY-fn}) 
if we simply investigate supersymmetric vacua.
The supersymmetric vacuum manifold
${\cal M}$ is defined by the vanishing potential energy density 
${\cal U}(\varphi) = 0$:
\begin{align*}
{\cal M} \ &:= \ 
\Big\{ (\varphi_a) \in {\mathbb C}^{n} \, \Big| \,
{\cal D} = F_a = {\cal U}_{\sigma} = 0 \Big\} \Big/U(1)
\; ,
\end{align*}
where $n$ is the number of scalar component fields in the GLSM.
The dividing $U(1)$ group indicates the abelian gauge symmetry.
Since we consider ${\cal N}=(2,2)$ supersymmetric theories, 
the manifold ${\cal M}$ becomes a K\"{a}hler manifold \cite{Zumino}
where the FI parameter denotes the scale of ${\cal M}$.

Under a generic configuration for chiral superfields $\Phi_a$ of charges $Q_a$,
the FI parameter $r$ receives a renormalization via wave-function
renormalizations of $\phi_a$. 
Thus the bare FI parameter $r_0$ is related to the renormalized one $r_R$
under the following equation:
\begin{align}
r_0 \ &= \ 
r_R + \sum_a Q_a 
\log \Big( \frac{\Lambda_{\rm UV}}{\mu} \Big) 
\; ,
\label{FI-renorm}
\end{align}
where $\Lambda_{\rm UV}$ and $\mu$ are the ultraviolet cut-off and 
the scale parameter, respectively. 
Thus we observe that the scale of ${\cal M}$ changes under the
renormalization group (RG) flow.
Studying the $\beta$-function of the FI parameter derived from (\ref{FI-renorm}),
we find whether
the effective theories expanded on ${\cal M}$ are asymptotically free
or not.
In particular, if we impose 
\begin{align}
\sum_a Q_a \ &= \ 0 \; , \label{CY-cond}
\end{align}
the FI parameter does not receive the renormalization.
Thus there appears a non-trivial conformal field theory 
in the IR limit.
{}From the geometric point of view,
the sum $\sum_a Q_a$ is equal to the first Chern
class $c_1({\cal M})$ of the vacuum manifold ${\cal M}$.
If the condition (\ref{CY-cond}) is satisfied on ${\cal M}$,
this manifold becomes a CY manifold.
Thus we refer (\ref{CY-cond}) to the ``CY condition.''
In attempt to study CY manifolds,
we impose this on the GLSM.

We usually study how massless effective theories 
are realized on the supersymmetric vacuum in ${\cal M}$.
Recall that in two-dimensional field theory
the ``massless'' modes are not well-defined because of 
the IR divergence in their two-point functions. 
The Coleman's theorem on 
non-existence of Nambu-Goldstone modes \cite{Coleman73} is
closely related to this difficulty.
In order to avoid this problem,
we assume that there exists an IR cut-off parameter.
Furthermore we take the large volume limit $r \to \infty$ 
when we consider a NLSM whose target space is the
vacuum manifold ${\cal M}$.
In this limit 
the FI parameter $r$ of GLSM can be related to the coupling constant
$g$ of the NLSM:
\begin{align*}
r \ &= \ \frac{1}{g^2} \; .
\end{align*}
This means that the large volume limit $r \to \infty$ is 
the weak coupling limit $g \to 0$.

Next we consider fluctuation fields around the vacuum. 
It is so complicated to analyze massless/massive fluctuation modes
that we perform here a general calculation.
Let us decompose scalar component fields 
$\varphi_{a}$
into three kinds of variables:
\begin{gather}
\varphi_a \ = \ \vev{\varphi_a} + \wt{\varphi}_a + \wh{\varphi}_a 
\; , \ls
\int \d^2 x \, \big( \wt{\varphi}_a + \wh{\varphi}_a \big) \ = \ 0 
\; ,
\label{decomp}
\end{gather}
where $\vev{\varphi_a}$, $\wt{\varphi}_a$ and $\wh{\varphi}_a$ mean
the vacuum expectation values (VEVs),
the fluctuation modes tangent and non-tangent to the vacuum manifold ${\cal M}$,
respectively.
They satisfy the following relations:
\bsubeq \label{cond-var} 
\begin{align}
\Scr{F}_{\alpha} (\varphi) \big|_{\rm VEV} 
\ &:= \ \Scr{F}_{\alpha} (\vev{\varphi})
\ \equiv \ 0 
\; , \label{cond-VEV} \\
\wt{\delta} \Scr{F}_{\alpha} (\varphi) \big|_{\rm VEV}
\ &:= \ \sum_a \wt{\varphi}_a \,
\frac{\del \Scr{F}_{\alpha} (\varphi)}{\del \varphi_a} 
 \Big|_{\varphi \equiv \vev{\varphi}}
\ \equiv \ 0 
\; , \label{cond-wt} \\
\wh{\delta} \Scr{F}_{\alpha} (\varphi) \big|_{\rm VEV}
\ &:= \ \sum_a \wh{\varphi}_a \, 
\frac{\del \Scr{F}_{\alpha} (\varphi)}{\del \varphi_a} 
 \Big|_{\varphi \equiv \vev{\varphi}}
\ \neq \ 0 
\; . \label{cond-wh} 
\end{align}
\esubeq
Note that $\Scr{F}_{\alpha} (\varphi)$ are the set of functions 
given by (\ref{D-cond}): 
$\Scr{F}_{\alpha} = \{ {\cal D}, F_a , {\cal U}_{\sigma} \}$.
The symbols $\wt{\delta}$ and $\wh{\delta}$ denote holomorphic
variations with respect to the complex variables $\varphi_a$.
Of course the VEVs $\vev{\varphi_a}$ satisfy the equation (\ref{cond-VEV}).
The equation (\ref{cond-wt}) provides that 
the first order variations of $\Scr{F}_{\alpha} (\varphi)$ with respect to the
fluctuation modes $\wt{\varphi}_a$ vanish.
This is nothing but the definition 
that $\wt{\varphi}_a$ move only tangent to the vacuum manifold ${\cal M}$.
The third equation (\ref{cond-wh}) means that the other fluctuation
modes $\wh{\varphi}_a$ do not propagate tangent to ${\cal M}$.
Substituting (\ref{decomp}) and (\ref{cond-var})
into the potential energy density ${\cal U}(\varphi)$ described by 
(\ref{SUSY-fn}),
we investigate the behaviors of the low energy effective theories.
If the equations (\ref{cond-VEV}) and 
(\ref{cond-wt}) furnish non-trivial relations among
the fluctuation modes $\wt{\varphi}_a$,
these modes constitute a supersymmetric NLSM whose target space is
${\cal M}$.
However, if these equations are trivially satisfied,
$\wt{\varphi}_a$ are free from constraints and propagate on
a flat space with potential energy. 
Then we find that
a field theory appears described by a superpotential of
fluctuation fields such as a LG superpotential $W_{\rm LG}$.

\subsection{Field configuration and supersymmetric vacuum manifold}

Now we are ready to analyze 
massless low energy effective theories in the
GLSM for ${\cal O}(-N+\ell)$ bundle on $\P{N-1}[\ell]$.
We consider a $U(1)$ gauge theory 
with $N+2$ chiral superfields $\Phi_a$ of charges $Q_a$. 
We set the field configuration to
\begin{gather}
\begin{array}{c|ccccc} \hline
\ \ \text{chiral superfield $\Phi_a$} \ \ 
& \ S_1 \ & \ \dots \ & \ S_N \ & \ P_1 \ & \ P_2 \ \\ \bcline{1-6}
\ \ \text{$U(1)$ charge $Q_a$} \ \ 
& \ 1 \ & \ \dots \ & \ 1 \ & \ -{\ell} \ & \ - N + {\ell} \ \\ \hline
\end{array}
\label{contents-O-CPNk}
\end{gather}
In addition we introduce a superpotential 
$W_{\rm GLSM} (\Phi) = P_1 \cdot G_{\ell} (S)$, where 
$G_{\ell} (S)$ is a function of chiral superfields $S_i$. 
This is a holomorphic homogeneous polynomial of degree ${\ell}$. 
Owing to the homogeneity, this polynomial has a following property: 
\begin{align}
\text{if} \ \ G_{\ell} (s) \ = \ \del_1 G_{\ell} (s) \ = \ \dots 
\ = \ \del_N G_{\ell} (s) \ = \ 0 
\ \ \ \to \ \ \ \text{then} \ \ \all s_i \ = \ 0 \; .
\label{qhomo-cond}
\end{align}
By definition, the numbers $N$ and $\ell$ are positive integers:
$\ell, N \in {\mathbb Z}_{>0}$.
We assume that these two integers satisfy
$1 \leq \ell \leq N-1$ and $2 \leq N$. 
The sum of all charges $Q_a$ vanishes (\ref{CY-cond})
in order to obtain non-trivial SCFTs on the CY manifold.


Now we consider the potential energy density and look for supersymmetric
vacua.
Imposing the Wess-Zumino gauge,
we write down the bosonic part of the potential 
energy density ${\cal U}(\varphi)$:
\bsubeq \label{pot-OCPNk}
\begin{align}
{\cal U} (\varphi) \ &= \ 
\frac{e^2}{2} {\cal D}^2  
+ \big| G_{\ell} (s) \big|^2 
+ \sum_{i=1}^N \big| p_1 \del_i G_{\ell} (s) \big|^2
+ {\cal U}_{\sigma} (\varphi)
\; , \label{pot-all} \\
{\cal D} \ &= \ 
r - \sum_{i=1}^N |s_i|^2 + {\ell} |p_1|^2 + (N-{\ell}) |p_2|^2
\; , \label{D} \\
{\cal U}_{\sigma} (\varphi)
\ &= \ 
2 |\sigma|^2 \Big\{ \sum_{i=1}^N |s_i|^2 + \ell^2 |p_1|^2 
+ (N-\ell) |p_2|^2 \Big\}
\; . \label{pot-sigma}
\end{align}
\esubeq
Imposing zero on them,
we obtain the supersymmetric vacuum manifold ${\cal M}$.
Since the Lagrangian has ${\cal N}=(2,2)$ supersymmetry
and the single $U(1)$ gauge symmetry, 
the vacuum manifold becomes a K\"{a}hler quotient space:
\begin{align}
{\cal M} \ &= \ 
\Big\{ (\varphi_a) \in {\mathbb C}^{N+3} \, \Big| \,
{\cal D} = G_{\ell} = p_1 \del_i G_{\ell} = {\cal U}_{\sigma} = 0 \Big\} \Big/ U(1)
\; ,
\label{def-vacua}
\end{align}
In attempt to study effective theories, 
we choose a point on ${\cal M}$ as a vacuum and give VEVs of scalar
component fields: $\varphi_a \equiv \vev{\varphi_a}$.
Then we expand the fluctuation modes around the vacuum.
In general, 
the structure of ${\cal M}$ is different for $r > 0$, $r =0$ and $r<0$ 
and there appear various phases in the GLSM.
The phase living in the $r >0$ region is referred to the ``CY phase,''
and the phase in $r <0$ is called to the ``orbifold phase.''
A singularity of the model emerges in the phase at $r=0$.
Thus we sometimes call this the ``singularity phase.'' 
We will treat these three cases separately.
We comment that in each phase
the vacuum manifold is reduced from the original ${\cal M}$. 
We often refer the reduced vacuum manifold to ${\cal M}_r \subset {\cal M}$. 
The VEVs of the respective phases can be set only in ${\cal M}_r$.

\subsection{Calabi-Yau phase} \label{CYphase}

In this subsection we analyze the CY phase $r >0$.
In this phase, ${\cal D} =0$ requires some $s_i$ cannot be zero and therefore
$\sigma$ must vanish.
If we assume $p_1 \neq 0$, 
the equations $G_{\ell} (s) = \del_i G_{\ell} (s) = 0$ 
with the condition (\ref{qhomo-cond}) imply that all $s_i$ must vanish.
However this is inconsistent with ${\cal D} = 0$.
Thus $p_1$ must be zero.
The variable $p_2$ is free as long as the condition ${\cal D}=0$ is satisfied.
Owing to these, 
the vacuum manifold ${\cal M}$ is reduced to ${\cal M}_{\rm CY}$
defined by
\begin{align}
{\cal M}_{\rm CY} \ &= \ 
\Big\{ (s_i; p_2) \in {\mathbb C}^{N+1} \, \Big| \,
{\cal D} = G_{\ell} (s) = 0 \, , \ r > 0
\Big\}
\Big/ U(1)
\; .
\label{def-vacua-CY}
\end{align}
Here we explain this manifold in detail.
This is an $(N-1)$-dimensional noncompact K\"{a}hler manifold.
The components $s_i$ denote the homogeneous coordinates of the complex
projective space $\P{N-1}$. 
The constraint $G_{\ell} (s) =0$ reduces 
$\P{N-1}$ to a degree $\ell$ hypersurface expressed to $\P{N-1}[\ell]$.
we find that $p_2$ is a fiber coordinate 
of the ${\cal O} (-N+\ell)$ bundle on $\P{N-1}[\ell]$.
Furthermore the vanishing sum of $U(1)$ charges indicates
that the FI parameter $r$ is not renormalized.
This is equivalent to $c_1({\cal M}_{\rm CY}) = 0$.
Thus we conclude that 
the reduced vacuum manifold ${\cal M}_{\rm CY}$ is nothing but a
noncompact CY manifold on which  
a non-trivial superconformal field theory is realized.

Let us consider a low energy effective theory.
We choose a vacuum and take a set of VEVs of the scalar component fields. 
Because $\exist \vev{s_i} \neq 0$, the $U(1)$ gauge symmetry
is spontaneously broken down completely.  
Next, we expand all fields in terms of fluctuation modes such as
$\varphi_a = \vev{\varphi_a} + \wt{\varphi}_a + \wh{\varphi}_a$.
We set $\vev{p_1}$, $\vev{\sigma}$ and $\wt{\sigma}$ to be zero.
Substituting them into the potential energy density 
(\ref{pot-OCPNk}), we obtain
\begin{align*}
{\cal U} \ &= \ 
\frac{e^2}{2} \Big\{
2 {\rm Re} \Big[ - \sum_{i=1}^N 2 \wh{s}_i \vev{\ol{s}_i} 
+ (N-\ell) \wh{p}{}_2 \vev{\ol{p}{}_2} \Big]
- \sum_{i=1}^N \big| \wt{s}_i + \wh{s}_i \big|^2
+ {\ell} |\wh{p}_1|^2 + (N-{\ell}) \big| \wt{p}_2 + \wh{p}_2 \big|^2 \Big\}^2
\\
\ & \ \ \ \ 
+ \Big| \sum_{i=1}^N \wh{s}_i \, \del_i G_{\ell} (\vev{s})
+ \sum_{k=2}^{\ell} \frac{1}{k!} \sum_{i_1, \cdots, i_k} 
(\wt{s} + \wh{s})_{i_1} \cdots (\wt{s} + \wh{s})_{i_k}
\cdot \del_{i_1} \cdots \del_{i_k} G_{\ell} (\vev{s}) \Big|^2
\\
\ & \ \ \ \ 
+ |\wh{p}_1|^2 \sum_{i=1}^N \Big|
\del_i G_{\ell} (\vev{s})
+ \sum_{k=2}^{\ell-1} \frac{1}{k!} \sum_{j_1, \cdots, j_k} 
(\wt{s} + \wh{s})_{j_1} \cdots (\wt{s} + \wh{s})_{j_k}
\cdot \del_i \del_{j_1} \cdots \del_{j_k} G_{\ell} (\vev{s}) \Big|^2
\\
\ & \ \ \ \ 
+ 2 |\wh{\sigma}|^2
\Big\{ \sum_{i=1}^N \big| \vev{s_i} + \wt{s}_i + \wh{s}_i \big|^2
+ {\ell}^2 |\wh{p}_1|^2 
+ (N-{\ell})^2 \big| \vev{p_2} + \wt{p}_2 + \wh{p}_2 \big|^2 \Big\}
\; . 
\end{align*}
Fluctuation modes $\wt{s}_i$ and $\wt{p}_2$ 
remain massless and 
move only tangent to ${\cal M}_{\rm CY}$
because they are subject to
$\wt{\delta} {\cal D} |_{\rm VEV} = \wt{\delta} G_{\ell} |_{\rm VEV} =0$.
The variation $\wt{\delta} (p_1 \del_i G_{\ell}) |_{\rm VEV} = 0$ 
indicates $\wt{p}_1 =0$.
The modes $\wh{\sigma}$, $\wh{p}_1$, $\wh{s}_i$ and $\wh{p}_2$ have mass 
$m^2 = {\cal O}(e^2 r)$.
The gauge field $v_m$ also acquires mass of order ${\cal O}(e^2 r)$ 
by the Higgs mechanism.
The fermionic superpartners behave in the same 
way as the scalar component fields  
because of preserving supersymmetry.
In the IR limit $e \to \infty$ and the large volume limit $r \to \infty$, 
the massive modes decouple from the system.
Thus we obtain 
\begin{align}
\text{${\cal N}=(2,2)$ 
supersymmetric NLSM on ${\cal M}_{\rm CY}$}
\label{sol-OCPNkr>k}
\end{align}
as a massless effective theory.
Notice that this description is only applicable in the large volume limit 
because the NLSM is well-defined in the weak coupling limit 
{}from the viewpoint of perturbation theory.
This effective theory becomes singular 
if we take the limit $r \to +0$
because the decoupled massive modes becomes massless.
This phenomenon also appears in the Seiberg-Witten theory 
\cite{SW9407, SW9408}, the black hole condensation \cite{Str9504, GMS9504}, and so on.

Let us make a comment on the target space ${\cal M}_{\rm CY}$.
By definition, the number $\ell$ means the degree of the vanishing polynomial
$G_{\ell} (s) = 0$, which gives a hypersurface in the projective space $\P{N-1}$.
We can see that if $\ell = 1$, $G_{\ell =1} (s) =0$ gives a linear constraint with
respect to the homogeneous coordinates $s_i$ and 
the hypersurface $\P{N-1}[\ell =1]$ is reduced to $(N-2)$-dimensional
projective space $\P{N-2}$.
This reduction does not occur if $2 \leq \ell \leq N-1$.
Here we summarize the shape of the target space ${\cal M}_{\rm CY}$ 
in Table \ref{classif-O-CPNk}:
\begin{table}[h]
\begin{center}
\begin{tabular}{c|c} \hline
 degree $\ell$ & vacuum manifold ${\cal M}_{\rm CY}$ \\ \hline
$\ell = 1$ & ${\cal O}(-N+1)$ bundle on $\P{N-2}$ \\
$2 \leq \ell \leq N-1$ & \ls ${\cal O}(-N+\ell)$ bundle on
$\P{N-1}[\ell]$ \ls 
\\ \hline
\end{tabular}
\end{center}
\caption{\sl Classification of ${\cal O}(-N+\ell)$ bundle on $\P{N-1}[\ell]$.}
\label{classif-O-CPNk}
\end{table}

\noi
Although the $\ell =1$ case has been already analyzed in the original paper
\cite{W93},
the other cases $2 \leq \ell \leq N-1$ are the new ones which have not
been analyzed.

\subsection{Orbifold phase} \label{orbifoldphase}

Here we consider the negative FI parameter region  $r <0$.
In this region
the total vacuum manifold (\ref{def-vacua}) 
is restricted to a subspace defined by
\begin{align}
{\cal M}_{\rm orbifold} \ &= \
\Big\{ (p_1, p_2; s_i) \in {\mathbb C}^{N+2} \, \Big| \,
{\cal D} = G_{\ell} = p_1 \del_i G_{\ell} = 0 \, , \ r < 0
\Big\} \Big/ U(1) 
\; . \label{vacua-OCPNkr<}
\end{align}
Since ${\cal D}=0$ does not permit $p_1$ and $p_2$ to vanish simultaneously, 
$\sigma$ must be zero. 
This subspace is quite different from ${\cal M}_{\rm CY}$ in the CY phase.
In addition, 
the shape of ${\cal M}_{\rm orbifold}$ is sensitive to the change of
the degree $\ell$ 
because of the existence of the constraints 
$G_{\ell} = p_1 \del_i G_{\ell} =0$ and the property (\ref{qhomo-cond}).
Thus let us analyze ${\cal M}_{\rm orbifold}$ and study massless
effective theories on it in the case of $3 \leq \ell \leq N-1$, 
$\ell =2$ and $\ell =1$, separately.

\subsection*{Effective theories of $\mbf{3 \leq  {\ell} \leq N-1}$}

Here we analyze the vacuum manifold ${\cal M}_{\rm orbifold}$ and 
massless effective theories of $3 \leq \ell \leq N-1$.
Owing to the constraints 
$G_{\ell} = p_1 \del_i G_{\ell} = 0$ and their property (\ref{qhomo-cond}), 
the manifold ${\cal M}_{\rm orbifold}$ is decomposed into the
following two subspaces: 
\bsubeq
\begin{align}
{\cal M}_{\rm orbifold} \big|_{3 \leq \ell \leq N-1}
\ &= \ {\cal M}_{r<0}^1 \cup {\cal M}_{r<0}^2
\; , \\
{\cal M}_{r <0}^{1} \ &:= \ 
\Big\{ (p_1 , p_2) \in {\mathbb C}^2 \, \Big| \, {\cal D} = 0 \, , \ r < 0
\Big\} \Big/ U(1)
\; , \label{re-vacua-3-1}
\\
{\cal M}_{r <0}^{2} \ &:= \ 
\Big\{ (p_2; s_i) \in {\mathbb C}^{N+1} \, \Big| \, {\cal D} = G_{\ell} = 0 \, , \ r < 0
\Big\} \Big/ U(1) \; .
\label{re-vacua-3-2}
\end{align}
\esubeq
In the former subspace the condition (\ref{qhomo-cond}) is trivially satisfied
whereas in the latter subspace it is satisfied non-trivially. 
Both of the two subspace include a specific region $p_1 = \all s_i = 0$.
The subspace ${\cal M}_{r<0}^1$ 
is defined as a one-dimensional 
{\sl weighted} projective space $\CP{1}{\ell, N-\ell}$ 
represented by two complex fields
$p_1$ and $p_2$ of $U(1)$ charges $-\ell$ and $-(N-\ell)$,
respectively.
The precise definition of the weighted projective space 
is in appendix \ref{wcp}.


Let us choose a supersymmetric vacuum and set VEVs of all scalar fields.
Then we expand all the fields around the VEVs.
Expanding the potential energy density (\ref{pot-OCPNk}) 
in terms of VEVs and fluctuation modes,
we obtain the following form:
\begin{align*}
{\cal U} \ &= \ 
\frac{e^2}{2} \Big\{
2 \, {\rm Re} \Big[ \ell \, \wh{p}_1 \vev{\ol{p}_1} + (N-\ell) \wh{p}_2
  \vev{\ol{p}_2} \Big]
- \sum_i \big| \wt{s}_i + \wh{s}_i \big|^2 
+ \ell \big| \wt{p}_1 + \wh{p}_1 \big|^2
+ (N-\ell) \big| \wt{p}_2 + \wh{p}_2 \big|^2 \Big\}^2
\\
\ & \ \ \ \ 
+ \big| G_{\ell} (\wt{s} + \wh{s}) \big|^2
+ \big| \vev{p_1} + \wt{p}_1 + \wh{p}_1 \big|^2
\cdot \sum_i \big| \del_i G_{\ell} (\wt{s} + \wh{s}) \big|^2
\\
\ & \ \ \ \ 
+ 2 |\wh{\sigma}|^2 \Big\{
\sum_i \big| \wt{s}_i + \wh{s}_i \big|^2
+ \ell^2 \big| \vev{p_1} + \wt{p}_1 + \wh{p}_1 \big|^2
+ (N-\ell)^2 \big| \vev{p_2} + \wt{p}_2 + \wh{p}_2 \big|^2
\Big\}
\; ,
\end{align*}
where $\vev{p_1}$ and $\vev{p_2}$ are VEVs of scalar components of
$P_1$ and $P_2$, respectively.
They live in the weighted projective space (\ref{re-vacua-3-1}).
Because the VEVs of $s_i$ are all zero, the $U(1)$ gauge symmetry 
is spontaneously broken to ${\mathbb Z}_{\alpha}$, where
$\alpha$ is the great common number between $\ell$ and $N-\ell$:
$\alpha = {\rm GCM}\{\ell, N-\ell\}$.
This potential energy density
provides that
all fluctuation modes $\wt{s}_i$ and $\wh{s}_i$ appear as linearly
combined forms such as $\wt{s}_i + \wh{s}_i$,
which do not acquire any mass terms.
The modes $\wt{p}_1$ and $\wt{p}_2$ remain massless and move tangent to
the subspace (\ref{re-vacua-3-1}). 
The other fluctuation modes acquire mass of order $m^2 = {\cal O}(e^2 |r|)$.
Thus, in the IR limit $e \to \infty$,
all the massive modes are decoupled from the system.
Thus we obtain the following massless effective theory:
\begin{gather}
\text{${\cal N}=(2,2)$ supersymmetric NLSM on $\CP{1}{\ell, N-\ell}$} 
\nn \\
\text{coupled to ``LG'' theory
with} \ \ 
\Big\{ W_{\rm LG} \ = \ (\vev{p_1} + P_1) G_{\ell} (S)
\Big\} \Big/ {\mathbb Z}_{\alpha}
\; ,
\label{sol-OCPNkr<k>3-2}
\end{gather}
where $P_1$ and $S_i$ are massless chiral superfields.
Note that the sigma model sector also contains the 
${\mathbb Z}_{\alpha}$ orbifold symmetry coming from the property of
$\CP{1}{\ell, N-\ell}$.
As is well known that the term $\vev{p_1} G_{\ell}(S)$ forms an ordinary LG
superpotential. 
Thus in the IR limit 
we can interpret that this term is marginal and flows to the ${\cal N}=(2,2)$
minimal model. 
The second term $P_1 \cdot G_{\ell} (S)$ is somewhat mysterious.
Since this term has not any isolated singularities
we might not obtain well-defined unitary CFT. 
This difficulty causes the noncompactness of the manifold 
${\cal M}_{\rm CY}$ which appears in the CY phase.

There are two specific points in the subspace $\CP{1}{\ell,N-\ell}$.
One is the point $p_2=0$ and the other is $p_1 =0$.
In the former point the gauge symmetry is enhanced to ${\mathbb Z}_{\ell}$.
Furthermore the mode $\wt{p}_1$ disappears and 
$\wh{p}_2$ becomes massless, 
which combines with a massless fluctuation
modes $\wt{p}_2$ linearly. 
This combined mode is free from any constraints.
The other massless modes $\wt{s}_i + \wh{s}_i$ in
(\ref{sol-OCPNkr<k>3-2})  remain massless and are also free from constraints.
Thus in the IR limit and the large volume limit,
the massless effective theory becomes an 
${\cal N}=(2,2)$ supersymmetric theory as
\begin{gather}
\Big\{ 
\text{CFT on ${\mathbb C}^1$} \ \otimes \ 
\text{LG theory
with $W_{\rm LG} = \vev{p_1} G_{\ell} (S)$}
\Big\} \Big/ {\mathbb Z}_{\ell}
\; . \label{sol-OCPNkr<k>3-1}
\end{gather}
This effective theory consists of $N+1$ massless chiral superfields such as 
$P_2$ and $S_i$, which live in the free and the LG sectors, respectively.
Since we take the IR limit,
this effective theory becomes an SCFT.
The LG sector flows to a well-known LG minimal model \cite{W93}.
Thus the sigma model sector is also 
a superconformal field theory.
Here we notice that
we did not integrate out but just decomposed all massive modes
in the above discussion 
because it is generally impossible to calculate the integration of them.
Thus the above effective theory is merely an approximate one.
If we will be able to integrate out all massive modes exactly,
the obtaining effective theory will be different from the above one.
In later section we will discuss the exact form of the effective theory. 

Next let us consider the latter point $p_1 = 0$ in the space $\CP{1}{\ell, N-\ell}$.
On this point the broken gauge symmetry is partially restored to ${\mathbb Z}_{N-\ell}$.
The massless fluctuation mode $\wt{p}_2$ becomes zero
whereas 
the massive mode $\wh{p}_1$ becomes massless, which combines with
$\wt{p}_1$ being free from any constraints.
Thus $P_1$ appears as a massless chiral superfield.
In the IR limit we obtain the supersymmetric massless effective theory 
such as
\begin{gather}
\Big\{
\text{LG theory
with $W_{\rm LG} = P_1 \cdot G_{\ell} (S)$ \ on 
${\mathbb C}^{N+1}$} 
\Big\} \Big/ {\mathbb Z}_{N-\ell}
\; ,
\label{sol-OCPNkr<k>3-3}
\end{gather}
which consists of $N+1$ massless chiral superfields such as $P_1$ and $S_i$.
This theory is not a well-defined LG theory because 
the superpotential $W_{\rm LG}$ has no isolated singularities.
We interpret the defect of isolated singularities 
as a noncompactness of the manifold ${\cal M}_{\rm CY}$ in the CY phase
via CY/LG correspondence (if this correspondence is satisfied in the
case of sigma models on noncompact CY manifolds.)
This property prevents from calculating a chiral ring of this model 
in the same way as unitary 
LG minimal models describing compact CY manifolds \cite{LVW89}.


Here we study massless effective theories on the subspace 
${\cal M}_{r<0}^2$ defined in (\ref{re-vacua-3-2}).
As mentioned before, 
there are non-trivial constraints in ${\cal M}_{r<0}^2$. 
Thus, as we shall see, the effective theories are also under these constraints.
In the same way as discussed before, 
we choose one point in the subspace ${\cal M}_{r<0}^2$
and make all the scalar fields fluctuate around it. 
Then we write down 
the expanded potential energy density (\ref{pot-OCPNk}) in terms of 
VEVs and fluctuation modes $\vev{\varphi_a}$, $\wt{\varphi}_a$ and $\wh{\varphi}_a$:
\begin{align*}
{\cal U} \ &= \ 
\frac{e^2}{2} \Big\{
2 \, {\rm Re} \Big[ 
- \sum_i \wh{s}_i \vev{\ol{s}_i} + (N-\ell) \wh{p}_2 \vev{\ol{p}_2} \Big]
- \sum_i \big| \wt{s}_i + \wh{s}_i \big|^2 
+ \ell \big| \wh{p}_1 \big|^2
+ (N-\ell) \big| \wt{p}_2 + \wh{p}_2 \big|^2 \Big\}^2
\\
\ & \ \ \ \ 
+ \Big| \sum_i \wh{s}_i \, \del_i G_{\ell} (\vev{s})
+ \sum_{k=2}^{\ell} \frac{1}{k!} \sum_{i_1, \dots, i_k} 
(\wt{s} + \wh{s})_{i_1} \cdots (\wt{s} + \wh{s})_{i_k} 
\cdot \del_{i_1} \cdots \del_{i_k} G_{\ell} (\vev{s}) \Big|^2
\\
\ & \ \ \ \ 
+ |\wh{p}_1|^2 \cdot \sum_i \Big| \del_i G_{\ell} (\vev{s})
+ 
\sum_{k=1}^{\ell-1} \frac{1}{k!} \sum_{j_1,\dots,j_k}
(\wt{s} + \wh{s})_{j_1} \cdots (\wt{s} + \wh{s})_{j_k}
\cdot \del_i \del_{j_1} \cdots \del_{j_k} G_{\ell} (\vev{s}) \Big|^2
\\
\ & \ \ \ \ 
+ 2 |\wh{\sigma}|^2 \Big\{
\sum_i \big| \vev{s_i} + \wt{s}_i + \wh{s}_i \big|^2
+ \ell^2 \big| \wh{p}_1 \big|^2 
+ (N-\ell)^2 \big| \vev{p_2} + \wt{p}_2 + \wh{p}_2 \big|^2 \Big\}
\; .
\end{align*}
This potential energy density indicates 
the following: The fluctuation modes 
$\wh{s}_i$, $\wh{p}_1$ and $\wh{p}_2$ are massive;
$\wt{s}_i$ and $\wt{p}_2$ move tangent to ${\cal M}_{r<0}^2$.
Thus, taking $e \to \infty$ and $|r| \to \infty$,
we obtain 
\begin{gather}
\text{${\cal N}=(2,2)$ supersymmetric NLSM on ${\cal M}_{r<0}^{2}$}
\; . \label{sol-OCPNkr<k>3-4}
\end{gather}
In this theory there exist $P_2$ and $S_i$ as massless chiral
superfields, which move tangent to ${\cal M}_{r<0}^2$.
Notice that in general points in ${\cal M}_{r<0}^2$ 
the $U(1)$ gauge symmetry is completely broken because of 
the existence of non-zero VEVs $\vev{s_i}$.
However, taking  $\all \vev{s_i} = \vev{p_1} = 0$ and
$\vev{p_2} \neq 0$ in the subspace ${\cal M}_{r<0}^2$, 
we find that the gauge symmetry is partially restored to
${\mathbb Z}_{N-\ell}$.


Even though the vacuum manifolds ${\cal M}_{r<0}^1$ and 
${\cal M}_{r<0}^2$ are connected on $p_1 = \all s_i = 0$, 
the effective theories given by
(\ref{sol-OCPNkr<k>3-3}) and (\ref{sol-OCPNkr<k>3-4})
are quite different from each other.
The reason is that
while the subspace ${\cal M}_{r<0}^1$ is free from
constraints $G_{\ell} = p_1 \del_i G_{\ell} =0$, 
in the subspace ${\cal M}_{r<0}^2$ these constraints are still valid
on the region $p_1 = \all s_i = 0$.
On account of the existence of these constraints, 
a phase transition 
occurs when the theory moves from one to the other.
Thus we conclude that a new phase appears on the subspace ${\cal M}_{r<0}^2$,
which has not been discovered 
in well-known GLSMs such as the models for ${\cal O}(-N)$
bundle on $\P{N-1}$, for $\P{N-1}[N]$, for resolved conifold, and so on.
We refer this phase to the ``3rd phase.''
Here we refer the phase on ${\cal M}_{r<0}^1$ to the orbifold phase,
as usual.

\subsection*{Effective theories of $\mbf{{\ell}=2}$}

Let us consider the orbifold phase of $\ell =2$.
In the same way as the previous analysis,
the constraints 
$G_{\ell} = p_1 \del_i G_{\ell} = 0$ and the property (\ref{qhomo-cond}) 
decompose the manifold ${\cal M}_{\rm orbifold}$ into two subspaces: 
\bsubeq
\begin{align}
{\cal M}_{\rm orbifold} \big|_{\ell =2} 
\ &= \ 
{\cal M}_{r<0}^1 \cup {\cal M}_{r<0}^2
\; , \\
{\cal M}_{r <0}^{1} \ &:= \ 
\Big\{ (p_1 , p_2) \in {\mathbb C}^2 \, \Big| \, {\cal D} = 0 \; , \ r < 0
\Big\} \Big/ U(1)
\ \equiv \ \CP{1}{2, N-2}
\; , \label{re-vacua-2-1}
\\
{\cal M}_{r <0}^{2} \ &:= \ 
\Big\{ (p_2; s_i) \in {\mathbb C}^{N+1} \, \Big| \, {\cal D} = G_{2} = 0 \; , \ r < 0
\Big\} \Big/ U(1)
\; . \label{re-vacua-2-2}
\end{align}
\esubeq
These two subspaces are glued in the region given by 
$p_1 = \all s_i = 0$.
Although this situation is same as to the case of $3 \leq \ell \leq N-1$,
the appearing massless effective theories are quite different. 


Here let us analyze the effective theories on the subspace 
${\cal M}_{r<0}^1 = \CP{1}{2,N-2}$.
We choose a point in this subspace  as a supersymmetric vacuum
and take VEVs of all scalar fields. 
Then we make all scalar fields fluctuate around the VEVs.
Fluctuation modes $\wt{p}_1$ and $\wt{p}_2$ are subject to the constraints 
such that they move only tangent to $\CP{1}{2, N-2}$.
The fluctuation modes $\wt{s}_i$ have no degrees of
freedom 
because of the variation of the constraint $p_1 \del_i G_2 =0$.
(In the case of (\ref{re-vacua-3-1}), 
the equations $G_{\ell} =0$ and $p_1 \del_i G_{\ell} =0$ 
are trivially satisfied in $\CP{1}{\ell,N-\ell}$.
These variations are also trivial.
However the case of $\ell =2$ is quite different.
By definition, some $\del_i \del_j G_2$ must have non-zero values. 
Thus even though the above equations 
are trivially satisfied in the subspace $\CP{1}{2,N-2}$, 
their variations give non-trivial constraints
on the fluctuation modes.)
Under these conditions we write down the potential energy density (\ref{pot-OCPNk})
in terms of VEVs $\vev{\varphi_a}$ and fluctuation modes
$\wt{\varphi}_a$ and $\wh{\varphi}_a$: 
\begin{align*}
{\cal U} \ &= \ 
\frac{e^2}{2} \Big\{
2 \, {\rm Re} \Big[2 \wh{p}_1 \vev{\ol{p}_1} + (N-2) \wh{p}_2
  \vev{\ol{p}_2} \Big]
- \sum_i \big| \wh{s}_i \big|^2 
+ 2 \big| \wt{p}_1 + \wh{p}_1 \big|^2
+ (N-2) \big| \wt{p}_2 + \wh{p}_2 \big|^2 \Big\}^2
\\
\ & \ \ \ \ 
+ \big| G_{2} (\wh{s}) \big|^2
+ \big| \vev{p_1} + \wt{p}_1 + \wh{p}_1 \big|^2
\cdot \sum_i \big| \del_i G_{2} (\wh{s}) \big|^2
\\
\ & \ \ \ \ 
+ 2 |\wh{\sigma}|^2 \Big\{
\sum_i \big| \wh{s}_i \big|^2
+ 4 \big| \vev{p_1} + \wt{p}_1 + \wh{p}_1 \big|^2
+ (N-2)^2 \big| \vev{p_2} + \wt{p}_2 + \wh{p}_2 \big|^2
\Big\}
\; .
\end{align*}
This function denotes the following:
The fluctuation modes $\wh{s}_i$, $\wh{p}_1$ and $\wh{p}_2$
acquire masses $m^2 = {\cal O}(e^2|r|)$;
the modes $\wt{p}_1$ and $\wt{p}_2$ remain massless and
move tangent to $\CP{1}{2,N-2}$.
Thus taking $e \to \infty$ and $|r| \to \infty$,
we obtain the massless effective theory described by
\begin{gather}
\text{${\cal N}=(2,2)$ supersymmetric NLSM on $\CP{1}{2, N-2}$}
\; .
\label{sol-OCPNkr<k=2-2}
\end{gather}
This sigma model has ${\mathbb Z}_{\alpha}$ orbifold symmetry 
coming from the property of $\CP{1}{2,N-2}$,
where $\alpha = {\rm GCM}\{2, N-2\}$.
This effective theory does not include massless LG theory.
The reason is that the degree two polynomial $G_2$ 
generates mass terms such as $|\vev{p_1}|^2 \sum_i |\del_i G_2|^2$.
(See, for example, \cite{Wa90}.)

Now we consider the effective theory on two specific points in $\CP{1}{2,N-2}$
like (\ref{sol-OCPNkr<k>3-1}) and (\ref{sol-OCPNkr<k>3-3}).
Expanding the theory on the one point $(p_1, p_2) = (p_1, 0)$,
the gauge symmetry is partially restored to ${\mathbb Z}_{2}$. 
Thus we obtain the effective theory on this specific point as 
\begin{gather}
\text{${\cal N}=(2,2)$ SCFT on ${\mathbb C}^1 / {\mathbb Z}_{2}$}
\; . \label{sol-OCPNkr<k=2-1}
\end{gather}
Note that this theory can possess the LG theory with a quadratic superpotential 
$W_{\rm LG} = \vev{p_1} G_{2} (S)$, which gives massive modes of $S_i$.  

The effective theory drastically changes if we expand the theory
on another point $(p_1, p_2) = (0,p_2)$ in $\CP{1}{2,N-2}$.
On this point, 
the broken gauge symmetry is enhanced to ${\mathbb Z}_{N-2}$ and 
the fluctuation modes $\wh{s}_i$ become massless with being free from any
constraints.
Both $\wt{p}_1$ and $\wh{p}_1$ are massless and linearly
combined in the potential energy density.
The remaining field $\wt{p}_2$ becomes zero 
because there exists a non-trivial variation
of the constraint ${\cal D} = 0$.
Summarizing these results,
we find that the following massless effective theory appears in the limit 
$e, |r| \to \infty$: 
\begin{gather}
\Big\{ 
\text{${\cal N}=(2,2)$ ``LG'' theory with $W_{\rm LG} = P_1 \cdot G_2 (S)$
\ on  ${\mathbb C}^{N+1}$}
\Big\} \Big/ {\mathbb Z}_{N-2}
\; .
\label{sol-OCPNkr<k=2-3}
\end{gather}
Although this superpotential also has no isolated singularities,
this theory should describe a non-trivial SCFT. 
We shall return here in later discussions.


We next study the massless effective theories 
on the subspace ${\cal M}_{r<0}^2$ defined in (\ref{re-vacua-2-2}).
The potential energy density is obtained as
\begin{align*}
{\cal U} \ &= \ 
\frac{e^2}{2} \Big\{
2 \, {\rm Re} \Big[ 
- \sum_i \wh{s}_i \vev{\ol{s}_i} + (N-2) \wh{p}_2 \vev{\ol{p}_2} \Big]
- \sum_i \big| \wt{s}_i + \wh{s}_i \big|^2 
+ 2 \big| \wh{p}_1 \big|^2
+ (N-2) \big| \wt{p}_2 + \wh{p}_2 \big|^2 \Big\}^2
\\
\ & \ \ \ \ 
+ \Big| \sum_i \wh{s}_i \, \del_i G_{2} (\vev{s})
+ \frac{1}{2!} \sum_{i,j} 
(\wt{s} + \wh{s})_{i} (\wt{s} + \wh{s})_{j} 
\cdot \del_{i} \del_{j} G_{2} (\vev{s}) \Big|^2
\\
\ & \ \ \ \ 
+ |\wh{p}_1|^2 \cdot \sum_i \Big| \del_i G_{2} (\vev{s})
+ \sum_{j}
(\wt{s} + \wh{s})_{j} \cdot \del_i \del_{j} G_{2} (\vev{s}) \Big|^2
\\
\ & \ \ \ \ 
+ 2 |\wh{\sigma}|^2 \Big\{
\sum_i \big| \vev{s_i} + \wt{s}_i + \wh{s}_i \big|^2
+ 4 \big| \wh{p}_1 \big|^2 
+ (N-2)^2 \big| \vev{p_2} + \wt{p}_2 + \wh{p}_2 \big|^2 \Big\}
\end{align*}
under the following constraints on fluctuation modes:
The fluctuations $\wt{s}_i$ and $\wt{p}_2$ move tangent to ${\cal M}_{r<0}^2$;
the other tangent mode $\wt{p}_1$ is zero;
the fluctuations $\wh{\varphi}_a$ are all massive of $m^2 = {\cal O}(e^2 |r|)$.
Thus the effective theory expanded around generic points in 
${\cal M}_{r<0}^2$ becomes  
\begin{gather}
\text{${\cal N}=(2,2)$ supersymmetric NLSM on ${\cal M}_{r<0}^{2}$}
\label{sol-OCPNkr<k=2-4}
\end{gather}
in the IR and large volume limit: $e, |r| \to \infty$.
The $U(1)$ gauge symmetry is completely broken if some $\vev{s_i} \neq 0$ exist.
On the other hand, if we expand the theory on a specific point
$p_1 = \all s_i = 0$,
the gauge symmetry is partially restored to ${\mathbb Z}_{N-2}$.


So far
we have studied the effective theories on all regions of the vacuum manifold
${\cal M}_{\rm orbifold}$ of $\ell =2$.
{}From the same reason discussed in the case of $3 \leq \ell \leq N-1$,
there exists a phase transition 
between the theories 
(\ref{sol-OCPNkr<k=2-3}) and (\ref{sol-OCPNkr<k=2-4}) 
because of the non-trivial constraint coming from the variation of
the equation $G_2 = 0$. 
Thus we find that the GLSM for the ${\cal O}(-N+2)$ bundle on
$\P{N-1}[2]$ also includes two phases in the negative FI parameter region.
The phase on (\ref{sol-OCPNkr<k=2-2}) is called the orbifold phase,
and we refer the phase on (\ref{sol-OCPNkr<k=2-4}) to the 3rd phase.

Here we illustrate the relation among the phases in the GLSM schematically 
in Figure \ref{solution-space3-fig}:
\begin{figure}[h]
\begin{center}
\hspace{1cm} \includegraphics[height=5.5cm]{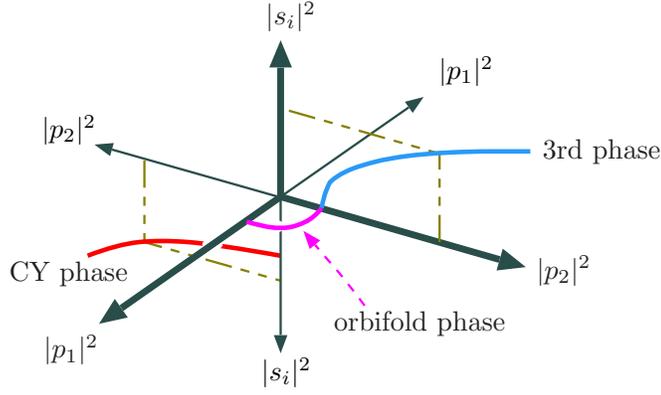}
\end{center}
\caption{\sl Various phases in GLSM for ${\cal O}(-N+\ell)$ bundle on
  $\P{N-1}[\ell]$ with $2 \leq \ell \leq N-1$.
The axes with thin/thick lines represent the vacuum space coordinates in the
positive/negative FI parameter regions, respectively.}
\label{solution-space3-fig}
\end{figure}

In the large volume limit $|r| \to \infty$, 
these three effective theories (\ref{sol-OCPNkr>k}),
(\ref{sol-OCPNkr<k>3-2}) and (\ref{sol-OCPNkr<k>3-4}) become
well-defined.
In later discussions we shall consider how these effective theories deform in
the small FI parameter limit $|r| \to 0$.
There we must consider the singular phase \cite{W93}.

\subsection*{Effective theory of $\mbf{{\ell}=1}$}

Finally we investigate the $\ell =1$ case.
Since the polynomial $G_{\ell =1} (S)$ is of degree one,
there exist some non-zero values of $\del_i G_{1} (S)$. 
Thus, combining this condition with the other constraints which define 
${\cal M}_{\rm orbifold}$, 
we find that $p_1$ must be zero 
and obtain the following reduced vacuum manifold:
\begin{align}
{\cal M}_{\rm orbifold} \big|_{\ell =1} \ &= \
\Big\{ (p_2; s_i) \in {\mathbb C}^{N+1} \, \Big| \, {\cal D} = G_{1} = 0 \; , \ r < 0
\Big\} \Big/ U(1)
\ =: \ {\cal M}_{r<0}^2 
\; .
\label{re-vacua-1-2}
\end{align}
Since this space is defined in the same way as (\ref{re-vacua-3-2})
and (\ref{re-vacua-2-2}),
we also referred it to ${\cal M}_{r<0}^2$.

After taking VEVs of scalar fields which live in (\ref{re-vacua-1-2}),
we make scalar fields fluctuate around the VEVs.
These fluctuation modes are subject to constraints:
$\wt{s}_i$ and $\wt{p}_2$ move only tangent to ${\cal M}_{r<0}^2$;
$\wt{p}_1$ is zero.
Substituting these into (\ref{pot-OCPNk}),
we obtain the expanded potential energy density 
\begin{align*}
{\cal U} \ &= \ 
\frac{e^2}{2} \Big\{ 
2 \, {\rm Re} \Big[ - \sum_i \wh{s}_i \vev{\ol{s}_i}
+ (N-1) \wh{p}_2 \vev{\ol{p}_2} \Big] 
- \sum_i \big| \wt{s}_i + \wh{s}_i \big|^2
+ \big| \wh{p}_1 \big|^2 + (N-1) \big| \wt{p}_2 + \wh{p}_2 \big|^2 \Big\}^2
\\
\ & \ \ \ \ 
+ \big| G_1 (\wh{s}) \big|^2 
+ \big| \wh{p}_1 \big|^2 \cdot \sum_i \big| \del_i G_1 (\vev{s}) \big|^2
\\
\ & \ \ \ \ 
+ 2 |\wh{\sigma}|^2 \Big\{
\sum_i \big| \vev{s_i} + \wt{s}_i + \wh{s}_i \big|^2
+ \big| \wh{p}_1 \big|^2 
+ (N-1) \big| \vev{p_2} + \wt{p}_2 + \wt{p}_2 \big|^2 \Big\}
\; . 
\end{align*}
This indicates that
the modes $\wt{s}_i$ and $\wt{p}_2$ remain massless
whereas 
the modes $\wh{s}_i$, $\wh{p}_1$ and $\wh{p}_2$ become massive. 
Thus the following massless effective theory appears in the limit $e, |r| \to \infty$:
\begin{gather}
\text{${\cal N}=(2,2)$ supersymmetric NLSM on ${\cal M}_{r<0}^2$}
\; ,
\label{sol-OCPNkr<k=1-2}
\end{gather}
where the $U(1)$ gauge symmetry is completely broken 
because $\exist \vev{s_i} \neq 0$.
While if we set the VEVs to $\all \vev{s_i} =0$,
the broken $U(1)$ gauge symmetry is enhanced to ${\mathbb Z}_{N-1}$. 
In addition 
the modes $\wh{s}_i$ become massless and are combined with 
the tangent modes $\wt{s}_i$, which are still under constraint $G_1 = 0$. 
The mode $\wt{p}_2$ becomes zero, 
which is derived from the variation of ${\cal D}=0$.
In this specific point 
we can see that the space ${\cal M}_{r<0}^2$ 
is deformed to ${\mathbb C}^{N-1}/{\mathbb Z}_{N-1}$
and the effective theory are
\begin{gather}
\text{${\cal N}=(2,2)$ SCFT on ${\mathbb C}^{N-1}/ {\mathbb Z}_{N-1}$}
\; .
\label{sol-OCPNkr<k=1-1}
\end{gather}
The two effective theories (\ref{sol-OCPNkr<k=1-2}) and (\ref{sol-OCPNkr<k=1-1}) 
are smoothly connected without 
any phase transitions coming from the variations of constraints.
Thus we find that 
in the $\ell =1$ case 
there exists only one phase in the negative FI parameter region.
We refer this to the orbifold phase, as usual.
Here we illustrate the schematic relation between the CY phase and the
orbifold phase in Figure 
\ref{solution-space1-fig}:
\begin{figure}[h]
\begin{center}
\hspace{1cm} \includegraphics[height=5.2cm]{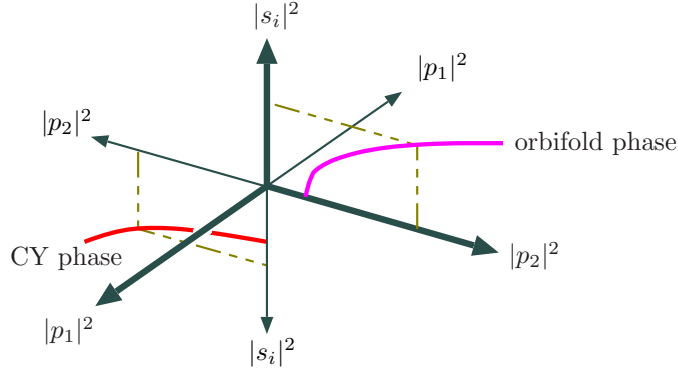}
\end{center}
\caption{\sl Various phases in GLSM for ${\cal O}(-N+1)$ bundle on
  $\P{N-1}[1]$.}
\label{solution-space1-fig}
\end{figure}

Note that the GLSM for ${\cal O}(-N+1)$ bundle on $\P{N-1}[1]$ is 
completely the same as the one for ${\cal O}(-N+1)$ bundle on $\P{N-2}$.
This is because the hypersurface $\P{N-1}[1]$ is nothing but
$\P{N-2}$.
Thus the vacuum structure and phases are also equal to each other.

\subsection{Singularity phase} \label{singularphase}

In this subsection let us analyze the singularity phase.
As mentioned before,
the effective theory (\ref{sol-OCPNkr>k}) becomes singular if $r \to +0$.
The effective theories in the orbifold and the 3rd phases 
also become singular if $r \to - 0$.
Thus we will study the singularity phase $r =0$ 
in order to avoid the singularities in effective theories.
In this analysis we will find that there appears a new branch.
Then we will discuss how to avoid the singularity. 

Here we study how the vacuum manifold (\ref{def-vacua}) is reduced 
in the $r =0$ phase.
If we assume $p_1 \neq 0$, then we obtain $\sum_i |s_i|^2 \neq 0$
from ${\cal D} =0$.
However the equations $G_{\ell} (s) =0$ and 
$p_1 \del_i G_{\ell} (s) =0$ insist that all $s_i$ vanish.
This is a contradiction.
Thus $p_1$ must be zero.
Under this condition,
we obtain two solutions.
One is obtained by ${\cal D}= 0$ and $\sigma =0$.
In general this solution has non-zero $\phi_a$, 
where $\phi_a$ are scalar component fields of chiral superfields.
The other solution is given by $\all \phi_a = 0$ and $\sigma$ is free.
We refer the former and latter solutions to the Higgs and Coulomb
branches, respectively.
These branches are similar to the ones of ${\cal N}=2$ SQCD in four
dimensions \cite{SW9408}.
(The CY, orbifold and 3rd phases are all in the Higgs branch.)
They are connected if all scalar component fields
vanish: $\all \varphi_a = 0$.
Now we analyze the effective theories on these two branches.

\subsection*{Higgs and Coulomb branches}

Let us consider the Higgs and Coulomb branches in detail.
In the Higgs branch,
there exist two supersymmetric vacuum solutions.
One is 
\bsubeq
\begin{align}
\all \vev{\phi_a} \ = \ {\cal O}(|r|) \ \to \ 0 
\; , \ls 
\vev{\sigma} \ = \ 0
\; .
\label{sing-sol-1}
\end{align}
This solution is smoothly connected with the supersymmetric vacuum
solutions in the phases of non-vanishing FI parameter.
The other is
\begin{align}
\vev{p_1} \ = \ 0 \; , \ls 
\all \vev{s_i} \, , \vev{p_2} \ : \ \ \ \text{arbitrary order}
\; , \ls
\vev{\sigma} \ = \ 0
\; ,
\label{sing-sol-2}
\end{align}
\esubeq
which is satisfied only on $r =0$.
Although the first solution (\ref{sing-sol-1}) appears in each GLSM,
the second solution (\ref{sing-sol-2}) does not satisfy the
supersymmetric vacuum condition
${\cal U} (\varphi) = 0$ in some GLSMs, for example, the GLSM for $\P{N-1}[N]$.

In the Coulomb branch,
we can set that $\all \vev{\phi_a} = 0$ and $\vev{\sigma}$ is free.
This solution appears only when the FI parameter vanishes.
If we choose $\vev{\sigma}$ to be zero, i.e., 
all the VEVs of scalar fields vanish $\vev{\varphi_a} =0$,
the Coulomb branch connects with the Higgs branch.  

Let us consider a massless effective theory in the Coulomb branch.
Since the scalar field $\sigma$ has mass dimension one,
we take the VEV $\vev{\sigma}$ to be very large.
Owing to this, 
all chiral superfields $S_i$, $P_1$ and $P_2$ acquire very large
masses via ${\cal U}_{\sigma} (\varphi)$ in (\ref{pot-OCPNk}).
Taking $\vev{\sigma} \to \infty$ and integrating out all massive
fields\footnote{Here we can integrate
  out $\Phi_a$ because the superpotential 
$W_{\rm GLSM}$ does not contribute to the deformation of the effective
  twisted superpotential $\wt{W}_{\rm eff} (\Sigma)$. 
For the precise derivation, see chapter 15 in \cite{Mirror03}.},
we obtain the following effective Lagrangian:
\begin{align*}
\Scr{L}_{\rm eff} \ &= \ 
\int \d^4 \theta \, \big\{ - {\cal K}_{\rm eff} (\Sigma, \ol{\Sigma}) \big\}
+ \Big( \frac{1}{\sqrt{2}} \int \d^2 \wt{\theta} \, \wt{W}_{\rm eff} (\Sigma)
+ c.c. \Big)
\; , \\
\wt{W}_{\rm eff} (\Sigma) \ &= \ 
- \Sigma \, t
- \sum_a Q_a \Sigma \Big\{ \log \Big( \frac{Q_a \Sigma}{\mu} \Big) - 1 \Big\}
\\
\ &= \ - \Sigma \Big(
t - \ell \log (- \ell) - (N-\ell) \log (- N+\ell) \Big)
\; .
\end{align*}
Note that the twisted superpotential was deformed by the quantum effects
coming from the integration of massive fields.
This effective Lagrangian presents the asymptotic form of 
the potential energy density 
\begin{align}
{\cal U}_{\rm eff} (\sigma) \ &= \ 
\frac{e_{\rm eff}^2}{2} \big| \del_{\sigma} \wt{W}_{\rm eff} (\sigma) \big|^2
\ = \ 
\frac{e_{\rm eff}^2}{2} 
\Big| t - {\ell} \log (-{\ell}) - (N-\ell) \log (-N+\ell) \Big|^2
\; .
\label{U-eff-in-C}
\end{align}
In order that the effective theory remains supersymmetric,
the potential energy density must be zero in a specific value of $\sigma$.
Notice that if the complexified FI parameter is given by
\begin{align}
t \ = \ {\ell} \log (-{\ell}) + (N-\ell) \log (-N+\ell)
\; ,
\label{singular-pt}
\end{align}
the potential energy density becomes always zero.
If it happens,
the effective theory does not have any mass gap and becomes singular
as a two-dimensional field theory.
Thus (\ref{singular-pt}) is the quantum singular point of the GLSM.
In the classical point of view, the value $t =0$ looks like a singular point
in the theory.
Integrating out the massive fields,
we find that the singular point moves to (\ref{singular-pt}).
The massless effective theories in Coulomb and Higgs branches 
are connected with each other avoiding this singular point.

\subsection*{CY/LG correspondence and topology change}

As mentioned before, the massless effective theories are only valid 
if we take the FI parameter to be infinitely large $|r| \to \infty$.
In this limit the effective theories are (partly) described by the
NLSMs.
However if we change the FI parameter to be small, 
the NLSM representations are no longer well-defined and 
must be deformed.
This phenomenon has been already studied in \cite{W93} as
following:
If the FI parameter goes to zero $r \to 0$,
the effective theory on the CY phase moves to the theory 
on the Coulomb branch in the singularity phase
avoiding the singular point.
Furthermore the effective theory connects to the LG theory in the
orbifold phase when $r \to - \infty$.

The above phenomenon suggests that,
rather than the LG theory being equivalent to the sigma model on the
CY manifold, they are two different phases of the same system, i.e., 
the system of the single GLSM.
Thus the CY/LG correspondence \cite{M89, V89, VW89, GVW89, M89-2, IV90, C91}
can be read from the phase transition.
In fact, it has been proved that 
the sigma model on $\P{4}[5]$
and the ${\mathbb Z}_5$-orbifolded LG theory with 
$W_{\rm LG} = G_5(S)$, 
which are equivalent to each other,
appear as the distinct phases in the single GLSM. 
Furthermore the topology change is also understood 
in the framework of the phase transition of the GLSM.
The flop of the resolved conifold ${\cal O}(-1) \oplus {\cal O}(-1) \to \P{1}$
is a typical example \cite{W93}.

Now let us apply the above discussions to the GLSM for 
${\cal O}(-N+\ell)$ bundle on $\P{N-1}[\ell]$.
For example we consider the relations among the various phases 
of $3 \leq \ell \leq N-1$,
where we have found three phases:
The CY phase on ${\cal M}_{\rm CY}$, 
the orbifold phase on ${\cal M}_{r<0}^1$ 
and the 3rd phase on ${\cal M}_{r<0}^2$.
Furthermore we have found four effective theories.
Let us discuss the relations among them:
\begin{itemize}
\item
The effective theory on the CY phase (\ref{sol-OCPNkr>k}) and 
the theory on the 3rd phase (\ref{sol-OCPNkr<k>3-4}) are related to
each other via a topology change, 
because the defining equations of the target spaces are equal except for the sign of the
FI parameter.
Furthermore these two effective theories are both sigma models, which
do not include the potential theory sectors such as a LG theory.
\item
(\ref{sol-OCPNkr<k>3-3}) and (\ref{sol-OCPNkr<k>3-4}) are connected
at the point $p_1 = \all s_i = 0$ in ${\cal M}_{\rm orbifold}$.
Since the former is the sigma model and the latter is the LG theory,
there exists a phase transition between these two theories,
which are equivalent to each other by the CY/LG correspondence.
\item 
Both the theories (\ref{sol-OCPNkr<k>3-1}) and (\ref{sol-OCPNkr<k>3-3})
are on the weighted projective space and are included 
in the theory (\ref{sol-OCPNkr<k>3-2}).
\item
The LG theory (\ref{sol-OCPNkr<k>3-1}) is equivalent to the
sigma model (\ref{sol-OCPNkr>k}) by the CY/LG correspondence.
\end{itemize}
Notice that the sigma model (\ref{sol-OCPNkr>k})
is not related to (\ref{sol-OCPNkr<k>3-3}) directly,
because 
the theory (\ref{sol-OCPNkr<k>3-3}) has
already been connected to (\ref{sol-OCPNkr<k>3-4})
while the CY/LG correspondence connects 
between two theories by one-to-one.
These connections are realized through the singularity phase.
Even though we wrote down the connections only from the qualitative point
of view, we can acquire non-trivial relations among the effective
theories as illustrated in Figure \ref{phases-OCPNk2}:
\begin{figure}[h]
\begin{center}
\includegraphics[height=6cm]{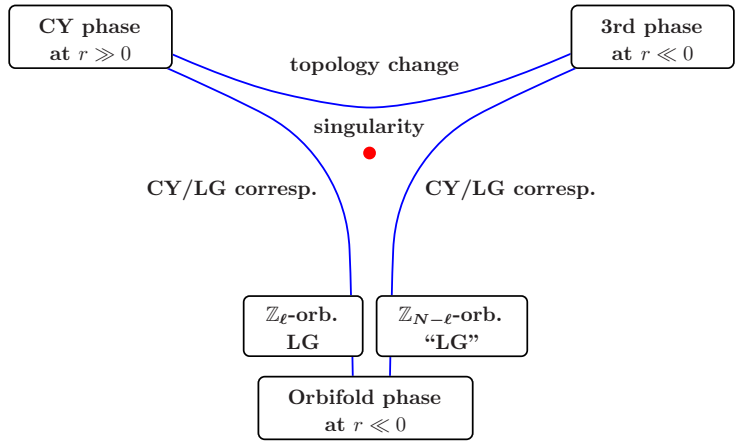}
\end{center}
\caption{\sl The relation among various phases around the singularity:
a conjecture.}
\label{phases-OCPNk2}
\end{figure}

In the case of compact CY manifolds,
we have already understood that 
the local rings in the LG theory are identified with the chiral rings 
of the SCFT and 
these chiral rings are related to the harmonic forms on such manifolds
\cite{LVW89}.
However we have no proof that this relation is also satisfied in the
case of noncompact CY manifolds.
Thus we must investigate the spectra of the above effective theories 
as a future problem.

As discussed before,
we have obtained various massless effective theories 
by decomposing all massive modes.
Thus they are just approximate descriptions
which must be deformed 
if we can exactly integrate out massive modes.
In the next section we will study the T-dual theory of the GLSM \cite{HV00}.
This formulation is so powerful to obtain the exact effective
theories.
Analyzing them exact theories we will re-investigate the massless
effective theories in the original GLSM.

\section{T-dual theory} \label{HV-O-CPNk}

In this section we consider T-dual of GLSMs.
It is quite significant to study it because 
we can obtain exact descriptions of the low energy effective
theories.
Furthermore 
they will also indicate how the exact effective theories are
realized in the original GLSM.
In fact, in the original GLSM,  
we obtained just approximate effective theories.
There we did not perform integrating-out but just decomposed all massive modes
because it is generally impossible to integrate them out. 
In the model proposed in \cite{HV00},
we calculate a function which is directly related to the
partition function.
Thus we will obtain exact effective theories as quantum field theory.

\subsection{General construction} \label{HV}

Here we briefly review the T-duality of a generic GLSM without any
superpotentials \cite{HV00}.
We start from the following Lagrangian in two-dimensional worldsheet:
\begin{align}
{\Scr{L}}' \ &= \ \int \! \d^4 \theta \, \Big\{ - \frac{1}{e^2}
\ol{\Sigma} \Sigma + \sum_a \Big( \e^{2 Q_a V + B_a} - \half (Y_a +
\ol{Y}{}_{\! a} ) B_a \Big) \Big\} 
+ \Big( \frac{1}{\sqrt{2}} \int \! \d^2 
\wt{\theta} \, (- \Sigma \, t ) + (c.c.) \Big) 
\; ,  \label{fund-L}
\end{align}
where $Y_a$ are twisted chiral superfields whose imaginary parts are periodic
of period $2 \pi$.
We incorporate real superfields $B_a$ as auxiliary fields.

Integrating out twisted chiral superfields $Y_a$,
we obtain $\ol{D}{}_+ D_- B_a = D_+ \ol{D}{}_- B_a = 0$,
whose solutions are written in terms of chiral superfields $\Psi_a$ and 
$\ol{\Psi}_a$ such as $B_a = \Psi_a + \ol{\Psi}_{\! a}$. 
When we substitute them into the Lagrangian (\ref{fund-L}), 
a GLSM Lagrangian appears:
\begin{align}
{\Scr{L}}' \Big|_{B_a = \Psi_a + \ol{\Psi}{}_a} 
\ &= \  \int \! \d^4 \theta \, \Big\{ 
- \frac{1}{e^2} \ol{\Sigma} \Sigma
+ \sum_a \ol{\Phi}{}_a \, \e^{2 Q_a V} \Phi_a \Big\} 
+ \Big( \frac{1}{\sqrt{2}} \int \! \d^2 \wt{\theta} \, (- \Sigma\, t) + (c.c.)
\Big) 
\nn \\
\ &\equiv \ 
\Scr{L}_{\rm GLSM}
\; , \label{GLSM-from-fund-L}
\end{align}
where we re-wrote $\Phi_a := \e^{\Psi_a}$.
On the other hand, 
when we first integrate out $B_a$ in the original Lagrangian (\ref{fund-L}),
we obtain 
\begin{align}
B_a \ = \ - 2 Q_a V + \log \Big( \frac{Y_a + \ol{Y}{}_{\! a}}{2} \Big) 
\; . \label{sol-B}
\end{align}
Let us insert these solutions into (\ref{fund-L}).
By using a deformation
\begin{align*}
\int \! \d^4 \theta \, Q_a V Y_a \ &= \ 
- \half Q_a \int \! \d^2 \wt{\theta} \, \ol{D}{}_+ D_- V Y_a 
\ = \ 
- \frac{1}{\sqrt{2}} Q_a \int \! \d^2 \wt{\theta} \, \Sigma Y_a
\; ,
\end{align*}
we find that a Lagrangian of twisted chiral superfields appears:
\bsubeq \label{T-dual-L}
\begin{gather}
\Scr{L}_{\rm T} \ = \ 
\int \! d^4 \theta \, \Big\{ - \frac{1}{e^2} \ol{\Sigma} \Sigma 
- \sum_a \Big( \half (Y_a + \ol{Y}{}_{\! a}) 
\log ({Y_a + \ol{Y}{}_{\! a}}) \Big) \Big\} 
+ \Big( \frac{1}{\sqrt{2}} \int \! \d^2 \wt{\theta} \, 
\wt{W} + (c.c.) \Big) \; , \\
\wt{W} \ = \ 
\Sigma \Big( \sum_a Q_a Y_a - t \Big) 
+ \mu \sum_a \, \e^{-Y_a} \; .
\label{general-T-W}
\end{gather}
\esubeq
This Lagrangian is T-dual of the gauge theory (\ref{GLSM-from-fund-L}).
Notice that
the twisted superpotential $\wt{W}$ is corrected by instanton
effects where the instantons are the vortices of the gauge theory.
In attempt to analyze a model satisfying $\sum_a Q_a = 0$,
the scale parameter $\mu$ is omitted by field re-definitions.
Relations between chiral superfields $\Phi_a$ in (\ref{GLSM-from-fund-L}) 
and twisted chiral superfields $Y_a$ in (\ref{T-dual-L}) are
\begin{align}
2 \ol{\Phi}{}_a \, \e^{2 Q_a V} \Phi_a \ &= \ Y_a + \ol{Y}{}_{\! a} \; .
\label{phi-y}
\end{align}
We can see that
the shift symmetry $Y_a \equiv Y_a + 2 \pi i$ comes from 
the $U(1)$ rotation symmetry on $\Phi_a$.
In the IR limit $e \to \infty$,
$\Sigma$ becomes non-dynamical and generates a following constraint from $\wt{W}$: 
\begin{align*}
\sum_a Q_a Y_a \ = \ t \; ,
\end{align*}
which corresponds to the condition ${\cal D} =0$ in the original GLSM.

In this formulation it is convenient to incorporate a function defined 
by
\begin{align}
\Pi \ &:= \ 
\int \d \Sigma \prod_a \d Y_a \, \exp \big( - \wt{W} \big) \; ,
\label{original-period}
\end{align}
where $\wt{W}$ is defined in (\ref{general-T-W}).
When we consider low energy effective theories of the theory
(\ref{T-dual-L}), 
we take the gauge coupling constant to be infinity $e \to \infty$.
In this limit $\Sigma$ is no longer dynamical and
becomes just an auxiliary field. 
Thus the function (\ref{original-period}) is re-written by
integrating-out of $\Sigma$:
\begin{align*}
\Pi \ &= \ 
\int \prod_a \d Y_a \, \delta \Big( \sum_a Q_a Y_a - t \Big)
\, \exp \Big( - \sum_a \e^{- Y_a} \Big)
\; .
\end{align*}
Via suitable field re-definitions,
we can read a LG theory of twisted chiral superfields. 
Moreover we also obtain a period integral of a ``mirror pair'' of the manifold 
which appeared in the effective theories in the original GLSM. 
Thus we often refer the function (\ref{original-period}) to the period integral.

Suppose the theory is topologically $A$-twisted \cite{W91}.
In the topologically $A$-twisted theory,
twisted chiral superfields are only valid while the other fields
such as chiral superfields and real superfields are all BRST-exact.
Due to this 
the Lagrangian is reduced only to the twisted superpotential
and the partition function is obtained as the integral of weight $\e^{- \wt{W}}$.
This is nothing but the period integral defined in
(\ref{original-period}).
Thus as far as considering the $A$-twisted sector,
the effective theories derived from this function are exact.

Unfortunately,
no one knows an exact formulation of a 
T-dual Lagrangian $\Scr{L}_{\rm T}$ 
of a GLSM with a generic superpotential $W_{\rm GLSM}$.
This is partly because 
the above formulation is only powerful 
when we consider T-duality of topologically $A$-twisted GLSMs.
As mentioned above, 
any deformations of $W_{\rm GLSM}$ are BRST-exact in the topological
$A$-twisted theory. 
However,
even though in the $A$-theories,
we can analyze T-dual theories of specific GLSMs with
superpotentials of type $W_{\rm GLSM} = P \cdot G_{\ell} (S)$, where
$G_{\ell} (S)$ is a homogeneous polynomial of degree $\ell$ with
respect to chiral superfields $S$. 
In order to do this,
let us deform the above period integral (\ref{original-period}) to
\begin{gather}
\wh{\Pi} 
\ = \ 
\int \d \Sigma \, \prod_a \d Y_a \, (\ell \Sigma) \, \exp \big( - \wt{W} \big) 
\; . \label{period}
\end{gather}
This function can be derived through the discussions of 
Cecotti and Vafa \cite{CV91}, and Morrison and Ronen Plesser
\cite{MP94}.
Here we omit a precise derivation.
Please see it in \cite{HV00}.
In this formulation we also take the IR limit $e \to \infty$
and integrate out the superfield $\Sigma$,
because we want to obtain the mirror dual
descriptions of the effective theories of the original GLSM with superpotential.
However the factor $\ell \Sigma$ in (\ref{period}) prevents from 
exact integrating-out of $\Sigma$.
Thus we need to replace this factor to other variable which does not
disturb the integration.
If we wish to obtain the LG description 
we replace the factor to the differential with respect to the FI parameter
such as $\ell \Sigma \to \ell \frac{\del}{\del t}$.
On the other hand when we derive the mirror geometry
we replace this to
the differential operator of an appropriate twisted chiral superfield derived from the 
negatively charged chiral superfield, for example, 
$\ell \Sigma \to \frac{\del}{\del Y_P}$, 
where $Y_P$ is the twisted chiral superfield 
of the chiral superfield $P$ of charge $-\ell$.
The resulting geometry has a ${\mathbb Z}_{\ell}$-type orbifold symmetry. 
For example,
we start from the GLSM for quintic hypersurface $\P{4}[5]$.
Performing T-duality and taking the IR limit,
we obtain not only the mirror dual geometry $\P{4}[5]/({\mathbb Z}_5)^3$
but also its LG description \cite{HV00}.
This procedure is so powerful that
we develop it in order to obtain the mirror descriptions of 
the noncompact CY manifolds. 
In subsections \ref{LG-twist-ONk-CPNk} and \ref{CY-twist-ONk-CPNk}
 we will study
how to obtain LG theories defined by the twisted superpotential 
and mirror dual geometries, respectively.
We can also obtain another geometry with a different orbifold symmetry
if we replace $\ell \Sigma$ to the differential of 
other suitable twisted chiral superfield.

\subsection{Field configuration}

Let us analyze the T-dual theory of the GLSM for the 
${\cal O}(-N+\ell)$ bundles on $\P{N-1}[\ell]$. 
The field configuration is assigned as follows: 
\begin{gather}
\begin{array}{c|ccccc} \hline
\ \ \text{chiral superfield $\Phi_a$} \ \ 
& \ S_1 \ & \ \dots \ & \ S_N \ & \ P_1 \ & \ P_2 \ \\ \bcline{1-6}
\ \ \text{$U(1)$ charge $Q_a$} \ \ 
& \ 1 \ & \ \dots \ & \ 1 \ & \ -{\ell} \ & \ - N + {\ell} \ \\ \bcline{1-6}
\ \ \text{twisted chiral superfield $Y_a$} \ \ 
& \ Y_1 \ & \ \dots \ & \ Y_N \ & \ Y_{P_1} \ & \ Y_{P_2} \ \\ \hline
\end{array}
\label{cont-TO-CPNk}
\end{gather}
The twisted chiral superfields $Y_a$ are periodic variables 
$Y_a \equiv Y_a + 2 \pi i$.
They are defined from the chiral
superfields $\Phi_a$ via (\ref{phi-y}).
As we have already discussed,
the twisted superpotential $\wt{W}$ and the period integral $\wh{\Pi}$,
given by the followings, play key roles:
\bsubeq \label{two-functions}
\begin{align}
\wt{W} \ &= \ 
\Sigma \Big( \sum_{i=1}^N Y_i - \ell Y_{P_1} - (N-\ell) Y_{P_2}
- t \Big) + \sum_{i=1}^N \e^{- Y_i} + \e^{- Y_{P_1}} + \e^{- Y_{P_2}}
\; , \label{original-wt-W} \\
\wh{\Pi} \ &= \
\int \d \Sigma \prod_{i=1}^N \d Y_i \, \d Y_{P_1} \, \d Y_{P_2} \,
({\ell} \Sigma) \,
\exp \big( - \wt{W} \big) 
\; .
\label{period-O-CPNk} 
\end{align}
\esubeq

Let us take the IR limit $e \to \infty$ in order to 
consider the low energy effective theories.
It is clear that the dynamics of $\Sigma$ is frozen and this
superfield becomes just an auxiliary superfield.
Thus we must replace the factor $\ell \Sigma$ in the period integral
(\ref{period-O-CPNk}) to appropriate variables.

\subsection{Mirror Landau-Ginzburg descriptions}
\label{LG-twist-ONk-CPNk}

In this subsection we will derive LG theories with orbifold symmetries.
In order to do this,
we change the variable $\ell \Sigma$ in the period integral
(\ref{period-O-CPNk}) to 
\begin{align*}
\ell \Sigma \ \to \ \ell \frac{\del}{\del t} 
\; .
\end{align*}
This replacing can be easily performed because of the existence of the
term $\Sigma (\sum_a Q_a Y_a - t)$ in $\wh{\Pi}$.
Then we integrate out the superfield $\Sigma$ and 
obtain 
\begin{align}
\wh{\Pi} \ &= \
{\ell} \frac{\del}{\del t} 
\int \prod_{i=1}^N \d Y_i \, \d Y_{P_1} \, \d Y_{P_2} \, 
\delta \Big( \sum_{i} Y_i - {\ell} Y_{P_1} - (N-{\ell}) Y_{P_2} - t \Big)
\,
\exp \Big( - \sum_{i} \e^{- Y_i} - \e^{- Y_{P_1}} - \e^{- Y_{P_2}} \Big)
\; .
\label{period1-On-CPnk}
\end{align}
Next let us solve the $\delta$-function in this function.
We note that there are two ways to solve it.
One is to write the variable $Y_{P_1}$ in terms of $Y_i$ and $Y_{P_2}$.
The other is to solve $Y_{P_2}$ by $Y_i$ and $Y_{P_1}$.
Both two solutions give consistent LG theories with orbifold symmetries.


\subsection*{Solution one: $\mbf{{\mathbb Z}_{\ell}}$ orbifolded LG theory}

Let us solve the variable $Y_{P_1}$ via the $\delta$-function 
in (\ref{period1-On-CPnk}):
\begin{align*}
- Y_{P_1} \ &= \ 
\frac{1}{{\ell}} \Big( t - \sum_{i=1}^N Y_i + (N-{\ell}) Y_{P_2} \Big)
\; .
\end{align*}
Performing the $t$-derivative in (\ref{period1-On-CPnk})
after the substitution of this solution, we obtain 
\begin{align*}
\wh{\Pi} \ &= \ 
\e^{t/\ell}
\int \prod_{i=1}^N \big( \e^{- \frac{1}{{\ell}} Y_i} \d Y_i \big) 
\, \big( \e^{\frac{N-{\ell}}{{\ell}} Y_{P_2}} \d Y_{P_2} \big) 
\, \exp \Big(
- \sum_{i} \e^{- Y_i} 
- \e^{t/\ell} \prod_{i} \e^{- \frac{1}{{\ell}} Y_i}
\e^{\frac{N-{\ell}}{{\ell}} Y_{P_2}} - \e^{- Y_{P_2}} \Big)
\; .
\end{align*}
It is clear that the integral measure is not canonical.
Transforming the variables into
$X_i^{\ell} := \e^{- Y_i}$ and 
$X_{P_2}^{\ell} := \e^{(N-{\ell}) Y_{P_2}}$, 
we obtain the following period integral with a canonical measure up to
an overall constant\footnote{It is not serious to ignore an overall constant.}:
\begin{gather}
\wh{\Pi} \ = \ 
\int \prod_{i=1}^N \d X_i \, \d X_{P_2} \, 
\exp \Big( - 
\sum_i X_i^{\ell} - X_{P_2}^{- \frac{{\ell}}{N-{\ell}}} 
- \e^{t/\ell} X_1 \cdots X_N X_{P_2} \Big) \; . 
\label{period-l-3}
\end{gather}
Since $Y_a$ are periodic with respect to the shifts of their imaginary parts
$Y_a \equiv Y_a + 2 \pi i$, 
the new variables $X_i$ and $X_{P_2}$ are symmetric under the
following phase shifts:
\begin{align}
X_i \ \mapsto \ \omega_i X_i \; , \ls
X_{P_2} \ \mapsto \ \omega_{P_2} X_{P_2} \; , \ls
\omega_i^{\ell} \ = \ 
\omega_{P_2}^{- \frac{{\ell}}{N-{\ell}}} 
\ = \ \omega_1 \omega_2 \cdots \omega_N \omega_{P_2} \ = \ 1 \; .
\label{phase-shift-l}
\end{align}
We can read from (\ref{period-l-3}) and (\ref{phase-shift-l}) that 
the following orbifolded LG theory appears:
\begin{align}
\Big\{ \wt{W}_{\ell} \ &= \ \sum_{i=1}^N X_i^{\ell} 
+ X_{P_2}^{- \frac{{\ell}}{N-{\ell}}} + \e^{t/\ell} X_1 \cdots X_N X_{P_2}
\Big\} \Big/ ({\mathbb Z}_{\ell})^{N}
\; .
\label{LG-twist-On-CPnk-ksol}
\end{align}
This theory is still ill-defined 
{}from the minimal model point of view.
Even though the terms of positive powers such as $X_i^{\ell}$ are
well-defined and they consist of ${\cal N}=2$ LG minimal model,
there exists a term $X_{P_2}^{- \frac{\ell}{N-\ell}}$, which does not generate
any critical points at finite $X_{P_2}$.
However there is an interpretation to avoid this difficulty.
Recall a discussion on the linear dilaton CFT and the Liouville theory \cite{GKP99, GK}.
(We prepare a brief review in appendix \ref{Giveon-Kutasov}.)
Based on this argument,
we can interpret the negative power term corresponds to
$Z_0^{-k}$ in (\ref{ill-W}), which gives an  
${\cal N}=2$ SCFT on the coset $SL(2,{\mathbb R})_k/U(1)$ at level $k$
assigned by 
\begin{align*}
k \ = \ \frac{\ell}{N-\ell} \; .
\end{align*}
This assignment is correct because the conformal weights $r_a$ in the
appendix \ref{Giveon-Kutasov} are all $r_a = 1/\ell$, where $n + 1= N$.
Thus we obtain $r_{\Omega} = \sum_a r_a - 1 = N/\ell - 1 \equiv 1/k$, 
which gives the above equation.
This theory is given as an ${\cal N}=2$ Kazama-Suzuki model on the
coset $SL(2,{\mathbb R})_k/U(1)$ at level $k$
\cite{KS}, which is the gauged WZW model 
on the two-dimensional Euclidean black hole \cite{W91WZW}.
Furthermore this theory is exactly equivalent to ${\cal N}=2$ Liouville theory 
of background charge $Q^2 = 2/k$ via T-duality \cite{GK, HK01}.
We will continue to argue in later discussions. 

\subsection*{Solution two: $\mbf{{\mathbb Z}_{N-\ell}}$ orbifolded LG theory}

In the same analogy of the previous discussion\footnote{{}From now on
  we omit overall constant factors which appear in the period integral.},
we study the theory of the second solution 
\begin{align*}
- Y_{P_2} \ &= \ 
\frac{1}{N-{\ell}} \Big( t - \sum_{i=1}^N Y_i + {\ell} Y_{P_1} \Big)
\;,
\end{align*}
which comes from the $\delta$-function in the period integral (\ref{period1-On-CPnk}).
Substituting this into (\ref{period1-On-CPnk}), we find 
\begin{align*}
\wh{\Pi} \ &= \ 
\int \prod_{i=1}^N \big( \e^{- \frac{1}{N-{\ell}} Y_i} \d Y_i \big) 
\, \big( \e^{\frac{{\ell}}{N-{\ell}} Y_{P_1}} \d Y_{P_1} \big) 
\, \exp \Big(
- \sum_{i} \e^{- Y_i} 
- \e^{\frac{t}{N-{\ell}}} \prod_{i} \e^{- \frac{1}{N-{\ell}} Y_i}
\e^{\frac{{\ell}}{N-{\ell}} Y_{P_1}} - \e^{- Y_{P_1}} \Big)
\; .
\end{align*}
Performing the re-definitions 
$X_i^{N-\ell} := \e^{- Y_i}$ and 
$X_{P_1}^{N-\ell} := \e^{\ell Y_{P_1}}$,
we find that the period integral has a canonical measure and 
the ``ill-defined'' LG theory with orbifold symmetry appears:
\begin{align}
\Big\{ 
\wt{W}_{N-{\ell}} \ = \ 
\sum_{i=1}^N X_i^{N-{\ell}} 
+ X_{P_1}^{- \frac{N-{\ell}}{{\ell}}} + \e^{\frac{t}{N-{\ell}}} X_1 \cdots X_N X_{P_1}
\Big\} \Big/ ({\mathbb Z}_{N-{\ell}})^{N}
\; .
\label{LG-twist-On-CPnk-nksol}
\end{align}
Applying the discussions in appendix \ref{Giveon-Kutasov} to the
negative power term in the superpotential $\wt{W}_{N-\ell}$,
we find that the theory is also described by 
the well-defined LG
theory with an orbifold symmetry coupled to 
${\cal N}=2$ Kazama-Suzuki model on the coset $SL(2,{\mathbb R})_k/U(1)$ at level
$k$, which is given by
\begin{align*}
\frac{N-\ell}{\ell} \ &= \ k \ = \ \frac{2}{Q^2} \; .
\end{align*}
where $Q$ is the charge of equivalent ${\cal N}=2$ Liouville theory.

\subsection{Mirror geometry descriptions} \label{CY-twist-ONk-CPNk}

In the previous subsection we found 
two orbifolded LG theories as exact effective theories.
They are obtained by solving the twisted chiral
superfields $Y_{P_1}$ and $Y_{P_2}$, respectively.
Next we will read geometric informations 
from the same period integral (\ref{period-O-CPNk}).
Here we will also obtain two solutions which are related to the LG theories.
The derivation procedure is so complicated that 
we try to imitate the method discussed in section 7.3 of \cite{HV00} 
and we develop detailed calculations, explicitly.
In order to obtain the geometric informations in the IR limit,
we integrate out the superfield $\Sigma$ in the period integral
(\ref{period-O-CPNk})
after the replacement of $\ell \Sigma$ in (\ref{period-O-CPNk})
to other variables, as we performed before.

\subsection*{$\mbf{{\mathbb Z}_{\ell}}$ orbifolded geometry}

Let us study how to obtain the geometry
with ${\mathbb Z}_{\ell}$-type orbifold symmetry.
Replacing $\ell \Sigma$ in the period integral (\ref{period-O-CPNk}) to 
\begin{gather*}
\ell \Sigma \ \to \ \frac{\del}{\del Y_{P_1}} 
\; , 
\end{gather*}
we can perform the integration of $\Sigma$ and obtain 
\begin{align}
\wh{\Pi} \ &= \ 
\int \prod_{i=1}^N \d Y_i \big( \e^{- Y_{P_1}} \d Y_{P_1} \big) 
\d Y_{P_2} 
\nn \\
\ & \LS \ls \times
\delta \Big( \sum_i Y_i - \ell Y_{P_1} - (N-\ell) Y_{P_2} - t \Big)
\exp \Big( - \sum_i \e^{- Y_i} - \e^{- Y_{P_1}} - \e^{- Y_{P_2}} \Big)
\; . \label{period-O-CPNk-l} 
\end{align}
We perform the re-definitions of the variables $Y_{i}$, $Y_{P_1}$ and $Y_{P_2}$:
\begin{align*}
\e^{- Y_{P_1}} \ &=: \ \wt{P}_1 \; , &
\e^{- Y_a} \ &=: \ \wt{P}_1 \, U_a 
& &\text{for $a = 1,\dots, {\ell}$}
\; , \\
\e^{- Y_{P_2}} \ &=: \ \wt{P}_2 \; , &
\e^{- Y_b} \ &=: \ \wt{P}_2 \, U_b & &\text{for $b = {\ell}+1,\dots, N$}
\; .
\end{align*} 
Substituting these re-defined variables into (\ref{period-O-CPNk-l}), 
we continue the calculation:
\begin{align}
\wh{\Pi} \ &= \ 
\int \prod_{i=1}^N \Big( \frac{\d U_i}{U_i} \Big)
\, \d \wt{P}_1 \, \Big( \frac{\d \wt{P}_2}{\wt{P}_2} \Big)
\, \delta \Big( \log \Big( \prod_{i} U_i \Big) + t \Big)
\, \exp \Big\{ - \wt{P}_1 \Big( \sum_{a=1}^{\ell} U_a + 1 \Big)
- \wt{P}_2 \Big( \sum_{b={\ell}+1}^N U_b + 1 \Big) \Big\}
\nn \\
\ &= \ 
\int \prod_{i} \Big( \frac{\d U_i}{U_i} \Big)
\, \d \wt{P}_2 \, \d u \, \d v
\, \delta \Big( \log \Big( \prod_{i} U_i \Big) + t \Big)
\, \delta \Big( \sum_{a} U_a + 1 \Big) 
\, \exp \Big\{ - \wt{P}_2 \Big( \sum_{b} U_b + 1 - uv \Big) \Big\}
\nn \\
\ &= \ 
\int \prod_{i} \Big( \frac{\d U_i}{U_i} \Big)
\, \d u \, \d v 
\, \delta \Big( \log \Big( \prod_{i} U_i \Big) + t \Big)
\, \delta \Big( \sum_{a} U_a + 1 \Big) 
\, \delta \Big( \sum_{b} U_b + 1 - uv \Big)
\; ,
\label{period-O-CPNk-homo1'}
\end{align}
where we introduced new variables $u$ and $v$ taking values in
 ${\mathbb C}$ and used a following equation 
\begin{align*}
\frac{1}{\wt{P}_2} \ &= \ 
\int \d u \, \d v \, \exp \big( \wt{P}_2 \, u v \big)
\; .
\end{align*}
It is obvious that the resulting function (\ref{period-O-CPNk-homo1'})
 still includes a non-canonical integral measure.
Thus we perform further re-definitions such as
\begin{align*}
U_a \ &=: \ \e^{- t/{\ell}} \frac{Z_a^{{\ell}}}{Z_1 \cdots Z_N} \; , 
\ls
U_b \ =: \ Z_b^{\ell} 
\; . 
\end{align*}
Note that the period integral (\ref{period-O-CPNk-homo1'}) is 
invariant under the following transformations acting on the new variables $Z_i$: 
\begin{gather*}
Z_a \ \mapsto \ \lambda \, \omega_a \, Z_a \; , \ \ \ 
Z_b \ \mapsto \ \omega_b \, Z_b 
\; , \ls 
\omega_a^{{\ell}} \ = \ \omega_b^{{\ell}} \ = \ \omega_1 \cdots \omega_N 
\ = \ 1 \; ,
\end{gather*}
where $\lambda$ is an arbitrary number taking in ${\mathbb C}^*$.
The $\omega_i$ come from the shift symmetry of the original variables
$Y_i \equiv Y_i + 2 \pi i$.
Combining these transformations we find that the period integral has 
${\mathbb C}^* \times ({\mathbb Z}_{\ell})^{N-2}$ symmetries.
Substituting $Z_i$ into (\ref{period-O-CPNk-homo1'}),
we obtain 
\begin{align*}
\wh{\Pi} \ &= \ 
\int \frac{1}{{\rm vol.}({\mathbb C}^*)} 
\prod_{i=1}^N \d Z_i \, \d u \, \d v 
\, \delta \Big(  \sum_{a=1}^{\ell} Z_a^{\ell} + \e^{t/\ell} Z_1 \cdots
Z_N \Big)
\, \delta \Big( \sum_{b={\ell}+1}^N Z_b^{\ell} + 1 - uv \Big) 
\; ,
\end{align*}
which indicates that the resulting mirror geometry is described by
\bsubeq \label{mirror-O-CPNk-homo2}
\begin{align}
&\hspace{-2.2cm}
\wt{\cal M}_{{\ell}} \ = \ 
\Big\{
(Z_i; u,v) \in {\mathbb C}^{N+2} \, \Big| \,
\big\{ {\cal F} (Z_i) \ = \ 0 \big\} \big/ {\mathbb C}^* \; , \ \ 
{\cal G} (Z_b; u,v) \ = \ 0
\Big\} \Big/ ({\mathbb Z_{{\ell}}})^{N-2} 
\; ,
\\
{\cal F} (Z_i) \ &:= \ \sum_{a=1}^{\ell} Z_a^{{\ell}} 
+ \psi Z_{1} \cdots Z_{\ell} 
\; , \ls
{\cal G} (Z_b; u, v) \ := \ \sum_{b={\ell}+1}^N Z_b^{{\ell}} + 1 - uv
\; , 
\label{eq-l-FG}
\\
\psi \ &:= \ \e^{t/\ell} Z_{\ell+1} \cdots Z_N \; .
\end{align}
\esubeq
This is an $(N-1)$-dimensional complex manifold.
It is guaranteed that $\wt{\cal M}_{\ell}$ is a CY
manifold 
because of the following reason:
We have already seen that 
the FI parameter $t$ in (\ref{period-O-CPNk}) does not
renormalized owing to the CY condition $\sum_a Q_a =0$,
which is also valid in the T-dual theory.
In addition, 
we took the IR limit $e \to \infty$ and obtained the above non-trivial
result.
This means that the sigma model on the above geometry is a
superconformal sigma model.

Let us study the manifold $\wt{\cal M}_{\ell}$ defined in (\ref{mirror-O-CPNk-homo2})
more in detail.
The equation ${\cal F}(Z_i) = 0$ denotes that the complex variables
$Z_a$ consist of the degree $\ell$ hypersurface in the projective space: 
$\P{\ell-1}[\ell]$.
This subspace itself is a compact CY manifold,
which is parametrized by a parameter $\psi$ which is subject to the equation
${\cal G} (Z_b;u,v) = 0$.
Moreover we can also interpret that 
the total space is a noncompact CY manifold
whose compact directions are described by $Z_i$, while
the variables
$u$ and $v$ run in the noncompact directions under the equations (\ref{eq-l-FG}).

Here let us comment on a relation between the manifold 
$\wt{\cal M}_{\ell}$ and the LG twisted superpotential (\ref{LG-twist-On-CPnk-ksol}).
As we have described in (\ref{mirror-O-CPNk-homo2}), 
$\wt{\cal M}_{\ell}$ has $({\mathbb Z}_{\ell})^{N-2}$ orbifold
symmetry, while
the LG theory (\ref{LG-twist-On-CPnk-ksol})
also holds this type of orbifold symmetry, i.e., 
the $({\mathbb Z}_{\ell})^{N}$ orbifold symmetry. 
When we combine the two equations in (\ref{eq-l-FG}) as follows:
\begin{align*}
F (Z_i, u,v) \ &\equiv \ {\cal F} (Z_i) + {\cal G} (Z_b, u,v)
\ = \  
\sum_{i=1}^N Z_i^{\ell} + \e^{t/\ell} Z_1 \cdots Z_N + (1 - uv)
\ = \ 0 \; . 
\end{align*}
This function $F = 0$ is quite similar to the LG
twisted superpotential $\wt{W}_{\ell}$ including
negative power term (\ref{LG-twist-On-CPnk-ksol}).
Recall that 
a LG theory written 
by a superpotential $W$ is identical with
a CY space defined by $W =0$ in a (weighted) projective space.
(See, for examples, \cite{M89, GVW89}.)
If we can apply this argument to the above result,
the LG theory (\ref{LG-twist-On-CPnk-ksol})
is identical with the sigma model on (\ref{mirror-O-CPNk-homo2}) 
and there also exists the CY/LG correspondence in the T-dual theory.

\subsection*{$\mbf{{\mathbb Z}_{N-\ell}}$ orbifolded geometry}

We have constructed the two LG theories: $({\mathbb Z}_{\ell})^{N}$ orbifolded LG
theory and $({\mathbb Z}_{N-\ell})^{N}$ orbifolded LG theory.
The former is related to the CY geometry $\wt{\cal M}_{\ell}$.
It is natural to consider there also exists 
a dual geometry related to the latter LG theory.
In the previous calculation, 
we replaced the $\ell \Sigma$
in the period integral (\ref{period-O-CPNk}) to 
$\frac{\del}{\del Y_{P_1}}$ and we obtained the 
$({\mathbb Z}_{\ell})^{N-2}$ orbifolded geometry.
Here we replace $\ell \Sigma$ to the differential with respect to 
$Y_{P_2}$, which is dual 
of the chiral superfield $P_2$ of charge $-(N-\ell)$:
\begin{gather*}
\ell \Sigma \ \to \ \frac{\ell}{N-\ell} \frac{\del}{\del Y_{P_2}} 
\; . 
\end{gather*}
Substituting this into (\ref{period-O-CPNk}),
we obtain the following expression:
\begin{align}
\wh{\Pi} \ &= \ 
\int \prod_{i=1}^N \d Y_i \, \d Y_{P_1} 
\big( \e^{- Y_{P_2}} \d Y_{P_2} \big) 
\nn \\
\ & \LS \ls \times
\delta \Big( \sum_i Y_i - \ell Y_{P_1} - (N-\ell) Y_{P_2} - t \Big)
\exp \Big( - \sum_i \e^{- Y_i} - \e^{- Y_{P_1}} - \e^{- Y_{P_2}} \Big)
\; . \label{period-O-CPNk-Nl} 
\end{align}
Let us perform the following re-definitions of 
the variables $Y_i$, $Y_{P_1}$ and $Y_{P_2}$:
\begin{align*}
\e^{- Y_{P_1}} \ &=: \ \wt{P}_1 \; , &
\e^{- Y_a} \ &=: \ \wt{P}_1 \, U_a & &\text{for $a = 1,\dots, {\ell}$}
\; , \\
\e^{- Y_{P_2}} \ &=: \ \wt{P}_2 \; , &
\e^{- Y_b} \ &=: \ \wt{P}_2 \, U_b & &\text{for $b = {\ell}+1,\dots, N$}
\; .
\end{align*} 
Substituting the re-defined variables into (\ref{period-O-CPNk-Nl})
and introducing auxiliary variables $u$ and $v$ in order to integrate
out $\wt{P}_1$ completely,
we obtain 
\begin{align}
\wh{\Pi} \ &= \ 
\int \prod_{i=1}^N \Big( \frac{\d U_i}{U_i} \Big)
\, \d u \, \d v 
\, \delta \Big( \log \Big( \prod_{i=1}^N U_i \Big) + t \Big)
\, \delta \Big( \sum_{b=\ell+1}^{N} U_b + 1 \Big) 
\, \delta \Big( \sum_{a=1}^{\ell} U_a + 1 - uv \Big)
\; .
\label{period-O-CPNk-homo1}
\end{align}
The integral measure still remains non-canonical. 
We next introduce further re-definitions of $U_i$:
\begin{align*}
U_a \ &=: \ Z_a^{N-{\ell}} \; , \LS
U_b \ =: \ \e^{- t/(N-{\ell})} \frac{Z_b^{N-{\ell}}}{Z_1 \cdots Z_N}
\; . 
\end{align*}
We can see that the map from $Z_i$ to $U_i$ is one-to-one modulo
the ${\mathbb C}^* \times ({\mathbb Z}_{N-\ell})^{N-2}$ action given by 
\begin{gather*}
Z_a \ \mapsto \ \omega_a Z_a \; , \ \ \ 
Z_b \ \mapsto \ \lambda \omega_b Z_b 
\; , \ls
\omega_a^{N-{\ell}} \ = \ \omega_b^{N-{\ell}} \ = \ \omega_1 \cdots \omega_N 
\ = \ 1 \; , 
\end{gather*}
where $\lambda$ takes value in ${\mathbb C}^*$.
On account of the above re-definitions and symmetries 
we find that the period integral is re-written as
\begin{align*}
\wh{\Pi} \ &= \ 
\int \frac{1}{{\rm vol.}({\mathbb C}^*)} 
\prod_{i=1}^N \d Z_i \, \d u \, \d v 
\, \delta \Big( \sum_{a=1}^{\ell} Z_a^{N-{\ell}} + 1 - uv \Big)
\, \delta \Big( \sum_{b=\ell+1}^N Z_b^{N-\ell} 
+ \e^{t/(N-\ell)} Z_1 \cdots Z_N \Big)
\; ,
\end{align*}
{}from which we can read the geometric information described by
\bsubeq \label{mirror-O-CPNk-homo1}
\begin{align}
&\hspace{-2.7cm}
\wt{\cal M}_{N-{\ell}} \ = \ 
\Big\{
(Z_i; u,v) \in {\mathbb C}^{N+2} \, \Big| \,
{\cal F}(Z_a; u,v) \ = \ 0 \; , \ \  \big\{ {\cal G}(Z_i) \ = \ 0 \big\}
\big/{\mathbb C}^* 
\Big\} \Big/ ({\mathbb Z_{N-{\ell}}})^{N-2}
\; , \\
{\cal F}(Z_a; u,v) \ &:= \ \sum_{a=1}^{\ell} Z_a^{N-{\ell}} + 1 - uv 
\; , \ls
{\cal G}(Z_i) \ := \ \sum_{b={\ell}+1}^N Z_b^{N-{\ell}} 
+ \psi Z_{\ell+1} \cdots Z_N 
\; , \\
\psi \ &:= \ \e^{t/(N-\ell)} Z_1 \cdots Z_{\ell}
\; .
\end{align}
\esubeq
This is also a noncompact CY manifold including 
a compact CY hypersurface $\P{N-\ell-1}[N-\ell]$, which is defined
by
${\cal G}(Z_i) =0$ and parametrized by $\psi$ with being subject to 
${\cal F}(Z_a;u,v) =0$.
Since the variables are the twisted chiral superfields,
we obtained the ${\cal N}=2$ supersymmetric NLSM on $\wt{\cal M}_{N-\ell}$
as a low energy effective theory of the T-dual theory.
We can see that the sigma model on this manifold is
identical with the LG theory described by (\ref{LG-twist-On-CPnk-nksol}).

\subsection{Return to the gauged linear sigma model}

As discussed before,
it has been proved that 
the ${\cal N}=2$ SCFT on coset $SL(2,{\mathbb R})_k/U(1)$ at level $k$
is exactly T-dual with the 
${\cal N}=2$ Liouville theory of background charge $Q$ under the
relation $Q^2 = 2/k$.
Let us apply this argument to the GLSM and its T-dual.
Notice that the massless effective theories in the T-dual theory are
exact, whereas the ones in the original GLSM are approximately realized.

Now let us recall that
if a CFT ${\cal C}$ has an abelian discrete symmetry group $\Gamma$,
the orbifold CFT ${\cal C}' = {\cal C}/\Gamma$ has a symmetry group
$\Gamma'$ which is isomorphic to $\Gamma$ and a new orbifold CFT
${\cal C}'/\Gamma'$ is identical to the original CFT ${\cal C}$.
Including this argument into the effective theories of the GLSM and its
T-dual theory of $2 \leq \ell \leq N-1$,
we find that the theories (\ref{sol-OCPNkr<k>3-1}) and (\ref{sol-OCPNkr<k>3-3})
are {\sl equivalent} to (\ref{LG-twist-On-CPnk-ksol}) and
(\ref{LG-twist-On-CPnk-nksol}), respectively.
Furthermore we can interpret that the theories (\ref{sol-OCPNkr<k>3-1}) and
(\ref{sol-OCPNkr<k>3-3}) are described by ${\cal N}=2$ Liouville
theories coupled to the well-defined LG minimal models 
as {\sl exact} effective theories.
As a result we obtain the non-trivial relations among the various
effective theories in the GLSM.
Here we refer one typical result.
The CY sigma model on (\ref{def-vacua-CY}) corresponds to 
(\ref{sol-OCPNkr<k>3-1}), which is deformed to the LG
theory coupled to the Liouville theory as an exact quantum theory.
This is equivalent to (\ref{LG-twist-On-CPnk-ksol}) via T-duality.
On account of the CY/LG correspondence, (\ref{LG-twist-On-CPnk-ksol}) and
sigma model on (\ref{mirror-O-CPNk-homo2}) are identical with each other.
Finally the original CY manifold (\ref{def-vacua-CY}) and
(\ref{mirror-O-CPNk-homo2}) are mirror dual with each other.
Notice that
the CY manifold ${\cal M}_{\rm CY}$ is also deformed 
because the Liouville theory indicates that the dilaton field propagates on the
target space \cite{NS0411, N0411}. 
Of course we find that there are the same relations among effective
theories (\ref{sol-OCPNkr<k>3-3}), (\ref{sol-OCPNkr<k>3-4}),
(\ref{LG-twist-On-CPnk-nksol}) and (\ref{mirror-O-CPNk-homo1}).

Let us consider the case $\ell =1$.
As discussed before, the GLSM has only two massless effective theories 
(\ref{sol-OCPNkr>k}) and (\ref{sol-OCPNkr<k=1-1}).
In addition, the subspace $\P{\ell-1}[\ell]$ in
(\ref{mirror-O-CPNk-homo2}) is ill-defined if $\ell =1$
and then the LG description (\ref{LG-twist-On-CPnk-ksol}) is also
ill-defined.
Thus the T-dual theory has only two descriptions 
(\ref{LG-twist-On-CPnk-nksol}) and (\ref{mirror-O-CPNk-homo1})
in the IR limit.
This situation is consistent with the result in \cite{HV00}, where 
the GLSM for ${\cal O}(-N)$ bundle on $\P{N-1}$ and its T-dual was discussed.

\section{Summary and Discussions} \label{summary}

We have studied the GLSM for noncompact
CY manifolds realized as a line bundle on a hypersurface
in a projective space.
This gauge theory has three non-trivial phases
and includes two types of four massless effective
theories in the IR limit.
Two theories are NLSMs on two distinct manifolds,
whereas the other two are LG theories coupled to 
complex one-dimensional SCFTs.
Following the conventional arguments,
we have interpreted that these four theories are related to each other under 
phase transitions such as
CY/LG correspondences and a topology change.
Performing the T-duality,
we have also obtained two types of four exact massless effective theories;
the two theories are the sigma models on newly appeared mirror CY manifolds, 
while the other two are the LG theories including the terms of negative power $-k$, 
which may be regarded as indicating ${\cal N}=2$ SCFTs 
on coset $SL(2,{\mathbb R})_k/U(1)$ at level $k$.
Since the SCFT on this coset is exactly 
equivalent to the ${\cal N}=2$ Liouville theory via T-duality,
we have argued that 
the LG effective theories derived from the original GLSM are exactly
realized by the Liouville theories coupled to the well-defined LG
minimal models.
The relations among the theories are illustrated in Figure \ref{relation-I-1}:
\begin{figure}[h]
\begin{center}
\includegraphics[width=0.75\textwidth]{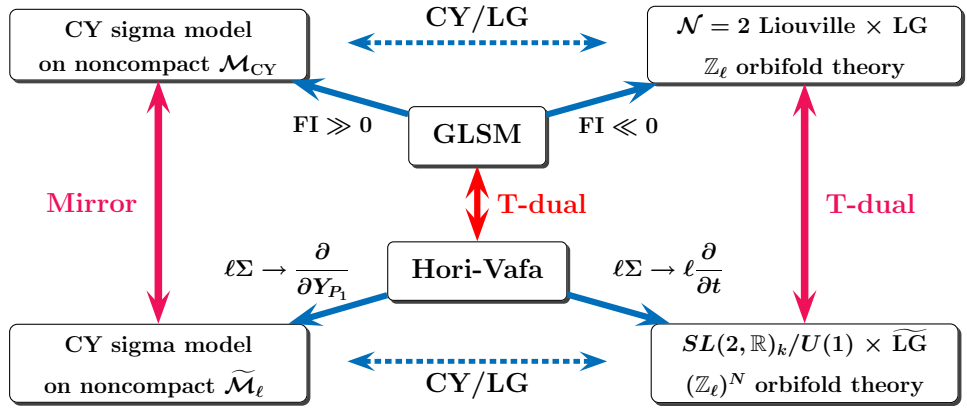}
\end{center}
\caption{\sl Relations among IR effective theories of GLSM and its T-dual.}
\label{relation-I-1}
\end{figure}

Utilizing the above relations,
we will obtain the topological charges of a CY manifold
from the exact effective theories in the T-dual theory,
even though we cannot directly calculate them in the
original sigma model.
Furthermore we will understand noncompact CY 
manifolds in detail from the mathematical point of view.
In addition,
we can interpret the holographic duality in type II string theory 
on noncompact (singular) CY manifolds
\cite{GK} as the phase transition and T-duality 
of the two-dimensional worldsheet theory
and will be able to understand this duality more closely
in the framework of the 
worldsheet sigma model description \cite{ES00, M0003, MMS, HK02}.


As mentioned in the introduction,
we have constructed the noncompact CY manifolds 
as line bundles on HSSs \cite{HKN0110}.
The base spaces HSSs can be seen as the submanifolds in the
projective spaces obtained by polynomials with additional symmetries \cite{HN99}:
the quadric surface $SO(N)/[SO(N-2) \times U(1)]$ is 
given by a polynomial of degree two with $SO(N)$ symmetry, and
$E_6/[SO(10) \times U(1)]$ has a set of differential
equations including $E_6$ isometry group.
These symmetries give the information of the complex structures of 
not only the base spaces HSSs but also the noncompact CY manifolds.
However, 
the T-dual theory \cite{HV00} is only valid when we consider the 
GLSM without a superpotential or with a superpotential given simply by
a homogeneous polynomial such as $W_{\rm GLSM} = P \cdot G_{\ell} (S)$.
Even though the polynomial $G_{\ell}(S)$ has an additional symmetry,
the period integral (\ref{original-period}) or (\ref{period}) 
{\sl cannot} recognize the existence of this additional symmetry.
Thus the T-dual theory does not map
all structures of the CY ${\cal M}$ to the mirror geometry completely.
For example,
we can argue the sigma model on the resolved conifold and
its mirror dual in the framework of GLSM and its T-dual,
however we have not even understood any correct descriptions for the deformed
conifold represented by the GLSM.
Therefore, if we wish to obtain the correct T-dual theories of the sigma models
on such noncompact CY manifolds,
we must improve the formulation 
so that it may recognize the complex structure of the manifold.
It is quite significant to solve this problem in order to understand mirror
symmetry for more general noncompact CY manifolds.

\section*{Acknowledgements}

The author would like to thank
Hiroyuki Fuji,
Kentaro Hori,
Takahiro Masuda,
Shun'ya Mizoguchi,
Toshio Nakatsu,
Kazutoshi Ohta,
Takuya Okuda,
Makoto Sakaguchi,
Hitoshi Sato,
Dan Tomino
and 
Atsushi Yamaguchi 
for valuable comments.
The author also thank Debashis Ghoshal for correspondence.
The author 
would also like to express the gratitude to 
Yukawa Institute for Theoretical Physics (YITP) for the hospitality
during the author's stay.
This work was supported in part by the JSPS Research Fellowships 
for Young Scientists (\#15-03926).

\begin{appendix}
\section*{Appendix}

\section{Conventions} \label{convention}

In this appendix we will write down the notation and convention 
which are modified from the ones defined in Wess-Bagger's book \cite{Wess-Bagger}.
In \cite{Wess-Bagger} supersymmetric field theory is defined in
four-dimensional spacetime.
However in this paper we discuss supersymmetric field theory in
two dimensions.
Thus let us first perform the dimensional reduction.
The coordinates in two dimensions $(x^0, x^1)$ are related to the four-dimensional
ones $(y^0, y^1, y^2, y^3)$:
\begin{align*}
(x^0,x^1) \ &\equiv \ (y^0, y^3) \; .
\end{align*}
Note that we perform the dimensional reduction for $y^1$- and $y^2$-directions.
Next we re-define the irreducible representation for spinors.
Weyl spinors $\psi_{\alpha}$ in four dimensions becomes Dirac spinors.
For convenience, we define the Dirac spinor indices in two dimensions
as \cite{W93}:
\begin{gather*}
(\psi{}^1,\psi{}^2) \ = \ (\psi{}^-,\psi{}^+) \; , \ls
(\psi_1,\psi_2) \ = \ (\psi_-,\psi_+) 
\;, \ls 
\psi{}^- \ = \ \psi_+ \; , \ls \psi{}^+ \ = \ -\psi_- 
\; , \\
\ve^{12} \ = \ \ve_{21} \ = \ 1
\ \ \ \to \ \ \ \ve^{-+} \ = \ \ve_{+-} \ = \ 1
\; , \\
(\psi^- , \psi^+)^{\dagger} 
\ = \ (\ol{\psi}{}^- , \ol{\psi}{}^+ ) 
\ = \ (\ol{\psi}{}_+, -\ol{\psi}{}_-)
\; .
\end{gather*}
Under the above convention,
the super differential operators $D_{\alpha}$ are also changed as follows:
\begin{align*}
D_{\pm} \ &= \ 
\frac{\del}{\del \theta^{\pm}} 
- i \ol{\theta}{}^{\pm} \big( \del_0 \pm \del_1 \big)
\; , \ls
\ol{D}{}_{\pm} \ = \ 
- \frac{\del}{\del \ol{\theta}{}^{\pm}} 
+ i \theta^{\pm} \big( \del_0 \pm \del_1 \big)
\; .
\end{align*}
Notice that the ordinary coordinate differentials are defined as
$\del_0 \equiv \del/\del x^0$ and $\del_1 \equiv \del/\del x^1$.
So far we wrote down the convention with respect to the
two-dimensional Minkowski spacetime.
When we consider a theory in two-dimensional Euclidean worldsheet,
we modify the coordinate $x^0$ to $x^0 = - i x^2$.

Under the above convention, 
let us briefly introduce the following irreducible superfields, i.e.,
{chiral superfields}, {vector (real) superfields} and 
{twisted chiral superfields}.

\noi
{\bf Chiral superfield}:
As in the case of four-dimensional theory,
a chiral superfield $\Phi$ is defined by $\ol{D}{}_{\pm}  \Phi = 0$.
We can expand a chiral superfield in terms of the fermionic
coordinates $\{ \theta^{\pm}, \ol{\theta}{}^{\pm} \}$ in the superspace:
\begin{align*}
\Phi(x,\theta, \ol{\theta}) 
\ &= \ 
\phi(x) + \sqrt{2} \theta{}^{+} \psi_{+} (x)
+ \sqrt{2} \theta{}^{-} \psi_{-} (x)
+ 2 \theta^+ \theta^- F(x) + \dots \; ,
\end{align*}
where $\phi (x)$ is a complex scalar field, 
$\psi_{\pm} (x)$ are Dirac spinors and 
$F (x)$ is a complex auxiliary field, whose mass dimensions are $0$, $1/2$
and $1$, respectively.
The part written by ``$+ \dots$'' involves only the derivatives of these
component fields $\phi$ and $\psi_{\pm}$.

\noi
{\bf Vector superfield}:
A vector superfield is defined by $V^{\dagger} = V$.
This setup is also same in four-dimensional spacetime.
We expand a vector superfield under the Wess-Zumino gauge:
\begin{align*}
V  (x, \theta, \ol{\theta})
\ &= \ 
\theta^+ \ol{\theta}{}^+ \big( v_0 + v_1 \big)
+ \theta^- \ol{\theta}{}^- \big( v_0 - v_1 \big)
- \sqrt{2} \theta^- \ol{\theta}{}^+ \sigma 
- \sqrt{2} \theta^+ \ol{\theta}{}^- \ol{\sigma}
\\
\ & \ \ \ \ 
- 2 i \, \theta^+ \theta^- 
\big( \ol{\theta}{}^+ \ol{\lambda}{}_+ + \ol{\theta}{}^- \ol{\lambda}{}_- \big)
+ 2 i \, \ol{\theta}{}^+ \ol{\theta}{}^- 
\big( \theta^+ \lambda_+ + \theta^- \lambda_- \big)
- 2 \theta^+ \theta^- \ol{\theta}{}^+ \ol{\theta}{}^- D \; .
\end{align*}
Note that we consider only $U(1)$ gauge theories in this paper,
where $v_0$ and $v_1$ are components of a $U(1)$ gauge potential, 
$\lambda_{\pm}$ are gaugino fields as Dirac spinors and
$D$ is a real auxiliary field.
The complex fields $\sigma$ and $\ol{\sigma}$ are coming from 
the dimensionally reduced components of four-dimensional $U(1)$ gauge potential.
In general we set this superfield to be dimensionless.

\noi
{\bf Twisted chiral superfield}:
A twisted chiral superfields is also an irreducible superfield in two dimensions.
The definition is $\ol{D}{}_+ Y = D_- Y = 0$.
Expanding a twisted chiral superfield $Y$ in terms of 
$\{ \theta^{\pm}, \ol{\theta}{}^{\pm} \}$,
we obtain 
\begin{align*}
Y (x, \theta, \ol{\theta}) \ &= \ y (x) + \sqrt{2} \theta^+ \ol{\chi}{}_+ (x)  
+ \sqrt{2} \ol{\theta}{}^- \chi_- (x) + 2 \theta^+ \ol{\theta}{}^- G
(x) + \dots \; .
\end{align*}
We denote a complex scalar field, Dirac spinors and an auxiliary
field as $y(x)$, $\{ \chi_-(x), \ol{\chi}{}_+ (x) \}$ and $G(x)$,
respectively.
The part ``$+ \dots$'' means derivatives of component fields $y(x)$,
$\chi_- (x)$ and $\ol{\chi}{}_+ (x)$.

We can construct a superfield $\Sigma$ for the field strength 
$F_{mn} \equiv \del_m v_n - \del_n v_m$ in the following way:
\begin{align*}
\Sigma \ := \ \frac{1}{\sqrt{2}} \ol{D}{}_+ D_- V 
\ &= \ \sigma - i \sqrt{2} \, \theta{}^+
\ol{\lambda}{}_+ -i \sqrt{2} \, \ol{\theta}{}^- \lambda_- 
+ \sqrt{2} \, \theta{}^+ \ol{\theta}{}^- (D-i F_{01}) 
\\
\ & \ \ \ \ 
-i \ol{\theta}{}^- \theta{}^- \, (\del_0 -\del_1) \, \sigma 
- i \theta{}^+ \ol{\theta}{}^+ \, (\del_0 + \del_1) \, \sigma 
+ \sqrt{2} \, \ol{\theta}{}^- \theta{}^+ \theta{}^- \,
(\del_0-\del_1) \, \ol{\lambda}{}_+ 
\\
\ & \ \ \ \ 
+\sqrt{2} \, \theta{}^+ \ol{\theta}{}^- \ol{\theta}{}^+ \,
(\del_0 + \del_1) \, \lambda_- 
- \theta{}^+ \ol{\theta}{}^- \theta{}^- \ol{\theta}{}^+ \,
(\del_0{}^2-\del_1{}^2) \, \sigma \; . 
\end{align*}
This is also a twisted chiral superfield $\ol{D}{}_+ \Sigma = D_- \Sigma = 0$.
This superfield is gauge invariant under the $U(1)$ gauge transformation.

Here let us define integral measures of the fermionic coordinates
$\theta^{\pm}$ and $\ol{\theta}{}^{\pm}$ in the superspace:
\begin{gather*}
\d^2 \theta \ := \ 
- \half \, \d \theta^+ \, \d \theta^- \; , \ls
\d^2 \wt{\theta} \ := \ 
- \half \, \d \theta^+ \, \d \ol{\theta}{}^- 
\; , \ls 
\d^4 \theta \ := \ 
- \frac{1}{4} \d \theta^+ \, \d \theta^- \, \d \ol{\theta}{}^+ \,
\d \ol{\theta}{}^- \; .
\end{gather*}
Thus the integral over $\theta^{\pm}$ and $\ol{\theta}{}^{\pm}$ are
obtained as follows:
\begin{align*}
\int \! \d^2 \theta \, \theta \theta \ &= \ 1 \; , \ls
\int \! \d^2 \wt{\theta} \, \theta^+ \ol{\theta}{}^- \ = \ \half 
\; .
\end{align*}
These definitions are slightly different from the ones 
in other papers, for example, \cite{HV00, Mirror03}.
We notice that we use the above convention in this paper.

\section{Weighted projective space} \label{wcp}

In this appendix we discuss a definition of one-dimensional weighted
projective space. 
The weighted projective space is slightly different from the 
(ordinary) projective space.
As we shall see,
the most significant difference is that
there exists an orbifold symmetry in the weighted projective space,
while the projective space does not have this symmetry.

First let us review one-dimensional (ordinary) projective space $\P{1}$. 
We prepare a two-dimensional complex plane without the origin
such as $W = {\mathbb C}^2 - \{ 0 \}$.
The coordinates on this plane are described as $(z_1, z_2)$.
The projective space $\P{1}$ is defined as a space whose coordinate is 
given by the ratio of the complex variables $z_1$ and $z_2$.
(The complex variables are called homogeneous coordinates in the
projective space.)
Under this definition, the two points $(z_1, z_2)$ and 
$(\lambda z_1, \lambda z_2)$ in $W$-plane are identified with each other:
\begin{align}
(z_1 , z_2) \ \simeq \ (\lambda z_1, \lambda z_2) \; ,
\label{id-cp}
\end{align}
where $\lambda$ is a variable of ${\mathbb C}^* = {\mathbb C} -
 \{0\}$. 
In other words,
all the points on the straight line through the origin of 
${\mathbb C}^2$ 
are identified via the above projection. 
Note that the projective space $\P{1}$ is diffeomorphic to the
two-sphere: $\P{1} \simeq S^2$.

Next we define a weighted projective space $\CP{1}{\ell,N-\ell}$.
Here we also prepare a two-dimensional complex plane $W$,
whose coordinates are expressed by $(z_1, z_2)$.
The weighted projective space is given as a space of 
complex coordinate defined by the ratio of $z_1$ and $z_2$ with
appropriate weights. 
The identification in the $W$-plane is the following:
\begin{align}
(z_1 , z_2) \ \simeq \ (\lambda^{\ell} z_1, \lambda^{N-\ell} z_2) \; ,
\label{id-wcp}
\end{align}
where both $\ell$ and $N-\ell$ are positive integers: 
$\ell, N-\ell \in {\mathbb Z}_{>0}$.
This identification has a residual symmetry with respect to the phases such as 
\begin{align}
(\omega^\ell z_1, \omega^{N-\ell} z_2) \ = \ (z_1, z_2) 
\; , 
\label{id-wcp-2}
\end{align}
where $\omega = \exp ( 2 \pi i \, \alpha )$ is the phase
of $\lambda$
and $\alpha$ is a great common number of $\ell$ and $N-\ell$ described by
$\alpha = {\rm GCM} \{ \ell, N -\ell \}$.
This does not exist in the definition of $\P{1}$.
In the case of $\P{1}$, the identification (\ref{id-cp}) fixes
the phase of homogeneous coordinates $z_1$ and $z_2$ completely.
On the other hand, the identification (\ref{id-wcp}) does not fix 
the phases of $z_1$ and $z_2$, and the residual symmetry (\ref{id-wcp-2}) exists.
Due to this, roughly speaking,
we can see that the weighted projective space is a projective space 
with ${\mathbb Z}_{\alpha}$ orbifold symmetry.
There are two specific points.
On the point $(z_1, z_2) = (z_1, 0)$, 
the orbifold symmetry is enhanced to ${\mathbb Z}_{\ell}$,
the other point $(z_1, z_2) = (0, z_2)$ generates ${\mathbb Z}_{N-\ell}$.
In addition,
if we choose $\ell = N-\ell$, the weighted projective space can be
reduced to the ordinary projective space
$\CP{1}{\ell, N-\ell=\ell} = \P{1}$, which has no longer an orbifold symmetry.

\section{Linear dilaton CFT and Liouville theory} \label{Giveon-Kutasov}

In this appendix we demonstrate the linear dilaton CFT and
the Liouville theory discussed in \cite{GKP99, GK}.
Let us consider the superstring propagating on the following
 the ten-dimensional spacetime: 
\begin{align*}
{\mathbb R}^{d-1,1} \times X^{2n}
\; , \ls
2 n  \ = \ 10 - d \; ,
\end{align*}
where $X^{2n}$ is a $2n$-dimensional singular CY manifold.
Sending the zero string coupling limit $g_s \to 0$ at fixed string
length $l_s$ gives rise to a $d$-dimensional theory without gravity describing
the dynamics of modes living near the singularity on $X^{2n}$.
This theory is holographic 
dual to string theory on a following background which approaches
at weak coupling region: 
\begin{align*}
{\mathbb R}^{d-1,1} \times {\mathbb R}_{\phi} \times {\cal M}
\ = \ 
{\mathbb R}^{d-1,1} \times {\mathbb R}_{\phi} \times S^1 \times {\cal M}/U(1)
\;, 
\end{align*} 
where ${\cal M}$ is a compact and non-singular manifold. 
The real line ${\mathbb R}_{\phi}$ is parameterized by $\phi$.

We can define CFT on each subspace.
On the flat space ${\mathbb R}^{d-1,1}$ we can define 
${\cal N}=1$ SCFT whose central charge is
\begin{align}
c_d \ &= \ \frac{3}{2} d \; .
\label{cc-d}
\end{align}
We describe the theory on ${\mathbb R}_{\phi}$ in terms of a linear dilaton
given by $\Phi = - \frac{Q}{2} \phi$.
The linear dilaton CFT has a central charge 
\begin{align}
c_{\phi} \ &= \ 1 + 3 Q^2 \; .
\label{cc-ld}
\end{align}
{}From the consistency of superstring propagation,
the worldsheet theory on ${\cal M}$ should be an ${\cal N}=1$ SCFT
with central charge $c_{\cal M} = 3 (n - 1/2 - Q^2)$. 
Moreover, if the manifold has a $U(1)$ symmetry, the theory on the coset manifold 
${\cal M}/U(1)$ must be an extended ${\cal N}=2$ SCFT with central
charge 
\begin{align}
c_{{\cal M}/U(1)} \ &= \ 3 (n - 1 - Q^2)
\; . 
\label{cc-coset}
\end{align}
Let us specialize the ${\cal N}=2$ SCFT on ${\cal M}/U(1)$ to the 
${\cal N}=2$ LG minimal model whose superpotential is defined 
in terms of $n+1$ chiral superfields $Z_a$:
\begin{align*}
W_{\rm LG} \ &= \ F (Z_a) \; , \ls 
F(\lambda^{r_a} Z_a) \ = \ \lambda F (Z_a) \; , \ls
a = 1,2, \dots, n+1 \; ,
\end{align*}
where $r_a$ are the conformal weights of the chiral superfields $Z_a$, respectively. 
Note that we have already understood properties of this minimal model
\cite{LVW89, Wa90}. 
The worldsheet central charge should
correspond to (\ref{cc-coset}) such as
\begin{align*}
c_{\rm LG} \ &= \ 3 \sum_{a=1}^{n+1} \big( 1 - 2 r_a \big)
\ \equiv \ 
c_{{\cal M}/U(1)} 
\; .
\end{align*}
If we introduce a new variable $r_{\Omega}$ 
with respect to the conformal weights $r_a$ such as 
$r_{\Omega} \equiv \sum_{a} r_a - 1$,
we can express the background charge $Q$ to $Q^2 = 2 r_{\Omega}$.

Here let us combine the above discussion with the conjectures 
proposed by Muhki and Vafa \cite{MV}, Ghoshal and Vafa \cite{GV95}, and 
Ooguri and Vafa \cite{OV95}, where 
they insisted that 
an ${\cal N}=2$ SCFT on the noncompact space 
${\mathbb R}_{\phi} \times {\cal M}$ can be given formally by the
``LG'' superpotential
\begin{align}
W \ &= \ - \mu Z_0^{-k} + F(Z_a)
\; ,
\label{ill-W}
\end{align}
where $Z_0$ is an additional chiral superfield and 
\begin{align}
k \ &= \ \frac{1}{r_{\Omega}} \ = \ \frac{2}{Q^2}
\; .
\label{k-Q}
\end{align}
This formulation is useful to describe the sigma model on deformed
conifold \cite{GV95}.
The first term in the superpotential appears to be ill-defined from
the LG minimal model point of view.
The corresponding potential does not have a minimum at the finite
value of $Z_0$.
The topological LG model with such a superpotential has already been
studied by Ghoshal and Mukhi \cite{GM93}, and Hanany, Oz and Ronen
Plesser \cite{HOP94}
in order to investigate two-dimensional string theory.
Moreover, in general, $k$ is not an integer, which makes (\ref{ill-W})
non-single valued.
Thus, it was proposed that this first term can be interpreted as an 
${\cal N}=2$ SCFT on the coset $SL(2,{\mathbb R})/U(1)$ at level $k$.
{}From the geometric point of view,
this coset space corresponds to a semi-infinite cigar, and 
in the IR limit this geometry deforms to the two-dimensional
Euclidean black hole \cite{W91WZW}.
This SCFT on the coset had been believed to be isomorphic to the 
Liouville theory in the sense of SCFT.
They are related by strong-weak coupling duality on the worldsheet:
The theory (\ref{ill-W}) can be valid as a Liouville theory 
in the large $Q$ limit, while 
this can be seen as a coset SCFT in the large $k$ limit ($k = 2/Q^2$).  
Finally, it has been proved that 
the ${\cal N}=2$ SCFT on the coset $SL(2,{\mathbb R})_k/U(1)$ is
exactly equivalent (or T-dual) to the ${\cal N}=2$ Liouville theory to each
other in any values of $k > 0$ \cite{HK01}. 
This equivalence was also proved by Tong in the framework of
two-dimensional domain wall physics in three-dimensional theory \cite{Tong03}.

To summarize, 
we find that the string theory on a singular CY manifold $X^{2n}$ can be
holographic dual to string theory as a product theory of 
the ${\cal N}=2$ SCFT on the coset $SL(2,{\mathbb R})_k/U(1)$
and the ${\cal N}=2$ LG minimal model on ${\cal M}/U(1)$.
The coset SCFT sector is also equivalent to 
the ${\cal N}=2$ Liouville theory on ${\mathbb R}_{\phi} \times S^1$.

\end{appendix}

}
\end{document}